\documentclass[a4paper,11pt]{article}
\pdfoutput=1
\usepackage{float}
\usepackage{jheppub}
\usepackage{amsmath,amsfonts,amssymb,amsthm,graphicx,color}
\usepackage{mathrsfs}
\usepackage[dvipsnames]{xcolor}
\usepackage{verbatim}
\usepackage{enumerate}
\usepackage{epsfig}
\usepackage{multirow}
\usepackage{comment}
\usepackage{datetime}
\usepackage{slashed}
\usepackage{framed}
\usepackage{longtable}
\usepackage[mathscr]{euscript}
\usepackage{cancel}
\usepackage{tensor}
\usepackage{mathtools}
\usepackage{caption}
\usepackage{subcaption}
\usepackage[multiple]{footmisc}
\usepackage{pgfplots}
\usepackage[T1]{fontenc}
\usepackage[utf8]{inputenc}
\usepackage{tikz}
\usepackage{placeins}
\usepackage{hyperref}
\usetikzlibrary{decorations.markings}
\usepackage{multirow}
\usepackage{graphicx}

\newcommand{\ft}[2]{{\textstyle\frac{#1}{#2}}}
\newcommand{\nn}{\nonumber}

\def\be{\begin{equation}}
	\def\ee{\end{equation}}
\def\bea{\begin{align}}
	\def\eea{\end{align}}
\def\beaq{\begin{eqnarray}}
	\def\eeaq{\end{eqnarray}}

\title{Resumming  Post-Minkowskian and Post-Newtonian gravitational waveform expansions}
\author[a]{Andrea Cipriani,}
\author[a]{Giorgio Di Russo,\footnote{Address after December 2024: School of Fundamental Physics and Mathematical Sciences, Hangzhou Institute for Advanced Study, UCAS, Hangzhou 310024, China}}
\author[a]{Francesco Fucito,}
\author[a]{Jose Francisco Morales,}
\author[b]{Hasmik Poghosyan }
\author[b]{and Rubik Poghossian}
\affiliation[a]{Dipartimento di Fisica, Universit\`a di Roma ``Tor Vergata" \& Sezione INFN  ``Roma Tor Vergata", Via della Ricerca Scientifica 1, 00133, Roma, Italy}
\affiliation[b]{Yerevan Physics Institute, Alikhanian Br. 2, 0036 Yerevan, Armenia}

\abstract{ We derive formulae that resum, at a given order in the soft limit, the infinite series of Post-Minkowkian (small gravitational coupling) or Post-Newtonian (small velocities) corrections to the  gravitational waveform  produced by particles moving along a general (open or closed) trajectory in Schwarzschild geometry in the probe limit. Specifying to the case of circular orbits,  we compute the wave form and the energy flux to order 30PN, and compare it against the available results in the literature.
The results are based on a novel hypergeometric representation of the solutions of the Heun equation (and its confluence), that leads to a simple mathematical proof of the Heun connection formula.

 }

\begin{document}
	\maketitle
	\flushbottom
	
	\section{Introduction}

 Recent advances in the sensitivity of experimental apparatus for detecting gravitational waves (GWs) necessitate increasingly precise theoretical computations. The next generation of space-based experiments, such as the Laser Interferometer Space Antenna (LISA), promises to revolutionize the GW detection by significantly enhancing sensitivity and exploring entirely new wavelength ranges compared to ground-based interferometers \cite{LISA:2022kgy}. LISA will detect long-wavelength GWs originating from sources like supermassive black hole mergers, extreme mass-ratio inspirals (EMRIs), galactic binaries, and the Stochastic Gravitational Wave Background—sources largely inaccessible to ground-based observatories.

The gravitational signals from these sources are diverse. For instance, GWs from supermassive black hole mergers will produce strong signals with large signal-to-noise ratios, while those from EMRIs, though weaker, will carry exceptionally detailed information about the gravitational sources.
Indeed,  EMRIs waveforms, spanning thousands of cycles during their inspiral phase, provide a unique opportunity to probe fundamental questions about the nature of gravitational sources, including their compactness, the presence or absence of horizons, light rings, echoes \cite{Cardoso:2019rvt},
and the validity of Einstein's theory of gravity and its possible extensions \cite{Gair:2012nm,Yunes:2013dva}.

Two main analytical tools have been developed for studying the problem we just described: the multipole expansion \cite{Thorne:1980ru,Damour:1982wm,Blanchet:2013haa} and perturbation theory \cite{Teukolsky:1973ha,Poisson:1993vp,Tagoshi:1993dm,Poisson:1994yf,Shibata:1994jx,Tagoshi:1994sm,Tanaka:1996lfd,Mano:1996vt,Mino:1997bx,Suzuki:1998vy,Fujita:2010xj,Fujita:2011zk}. The latter, valid in the probe limit of the binary, the so called EMRI systems, is the focus of this paper. In this framework, GWs are computed by solving the Einstein equations to linear order in the perturbations induced by the motion of a lighter body orbiting around a very massive black hole (BH). The radial and angular propagations of the waves are governed by ordinary differential equations of the confluent Heun type. Analytic solutions of these  equations are not known, but it is possible to write them as an expansion in the Newton constant  (Post-Minkowskian, or PM, approximation) or in the orbital velocity (Post-Newtonian, or PN, approximation).

 In the case of circular orbits, the PM and PN approximations are linked and characterised by a single expansion parameter $v$ (the tangential velocity).
 The current state-of-the-art for the emitted energy in perturbation theory  is the 22nd PN (order $v^{44}$) computation in \cite{Fujita:2012cm}. The expansion  is known to be converging slowly \cite{Cutler:1993vq,Poisson:1993vp} due to the presence of logarithmic terms, leading to an infinite tower of tails and tails of tails non-local interactions    \cite{Damour:2008gu,Damour:2009kr,Buonanno:2009zt,Damour:2012ky}.
In \cite{Aminov:2020yma,Bianchi:2021xpr,Bianchi:2021mft,Bonelli:2021uvf,Bonelli:2022ten,Consoli:2022eey,Bautista:2023sdf,Aminov:2023jve,ToVSapQNM,Fucito:2023afe} novel techniques, capitalizing on recent advances in quantum field theories, have been exploited to  derive an exact formula for the  Heun connection  matrix and alternative formulations of the waveforms were proposed in terms of the instanton partition function of a SU(2)$^2$ quiver gauge theory. Following this path the infinite tower of logarithmic terms have been resummed into exponentials.

More recently, multipole methods \cite{Bini:2023fiz,Bini:2024ijq,Bini:2024rsy}, scattering amplitude techniques \cite{DiVecchia:2023frv,Brandhuber:2023hhy,Herderschee:2023fxh,Elkhidir:2023dco,Georgoudis:2023lgf,DeAngelis:2023lvf,Brandhuber:2023hhl,Georgoudis:2023eke,Georgoudis:2023ozp,Georgoudis:2024pdz,Brunello:2024ibk,Alessio:2024onn} and black hole perturbation theory \cite{Fucito:2024wlg} have been applied to the study of gravitational scattering processes, fostering a fruitful interplay and mutual cross-checking among these complementary approaches. Within the quantum field theory framework, the PM series describes the expansion in loops (weak coupling) of the scattering amplitude keeping finite the relative velocity of the binary.
Using perturbation theory, we will show that at any fixed order in the gravitational coupling, the entire series of PN corrections to the probe limit of the gravitational waveform can be  resummed and expressed in terms of a single hypergeometric function. Analogous results hold for the PN expansion itself, where the exact dependence on the  gravitational coupling corrections  is determined at any given order in the small-velocity limit.

The PM and PN series provide novel hypergeometric representations of confluent Heun function  closely in spirit to the well studied incoming and upgoing MST solutions \cite{Mano:1996vt}. The crucial difference between the two, is that while the MST ansatz involves an infinite sum of hypergeometric functions, our  solution is written in terms of a single hypergeometric function and its first derivative. The crucial advantage of this representation is that it provides a simple mathematical proof of the Heun connection formulae originally found in \cite{Aminov:2020yma,Bianchi:2021xpr, Bonelli:2021uvf,Bonelli:2022ten}. The connection matrix coefficients are given as ratio of gamma functions with arguments depending on a single
(but highly non-trivial) function codifying the holonomies of the confluent Heun function around the singular point.  This function, known as the "renormalized angular momentum" in the MST context \cite{Tagoshi:1993dm,Poisson:1994yf,Shibata:1994jx,Tagoshi:1994sm,Tanaka:1996lfd,Mano:1996vt,Mino:1997bx,Suzuki:1998vy,Fujita:2010xj,Fujita:2011zk},
     can be related to the so called Seiberg-Witten period of a ${\cal N}=2$ supersymmetric gauge theory \cite{Consoli:2022eey,Bautista:2023sdf,Aminov:2023jve,ToVSapQNM,Fucito:2023afe} or
  the "anomalous dimension" in the effective field theory description of the gravitational wave \cite{Goldberger:2009qd,Trestini:2023wwg,Almeida:2021jyt,Edison:2023qvg}.
    The Heun connection formula has been recently applied in a variety of physical contexts, including the study of QNMs for gravity backgrounds (BHs, fuzzballs, D-branes), correlators in thermal field theories, holography, cosmological perturbations and scattering amplitudes in gravity \cite{Bianchi:2021xpr,Bianchi:2021mft,Fioravanti:2021dce,Bianchi:2022wku,Dodelson:2022eiz,Dodelson:2022yvn,Imaizumi:2022qbi,Fioravanti:2022bqf,Imaizumi:2022dgj,Bianchi:2022qph,Dodelson:2023vrw,Bianchi:2023rlt,Bianchi:2023sfs,Giusto:2023awo,Aminov:2023jve,BarraganAmado:2023apy,Lei:2023mqx,BarraganAmado:2023wxt,Fucito:2023afe,Bautista:2023sdf,Ge:2024jdx,Cipriani:2024ygw,Arnaudo:2024rhv,Bianchi:2024vmi,Bianchi:2024mlq,DiRusso:2024hmd,Bianchi:2024rod,Aminov:2024aan,BarraganAmado:2024tfu,Bautista:2024emt,Liu:2024eut,Silva:2025khf,Aprile:2025hlt,Bhatta:2025kil}.

    To keep the material self-contained, in this paper we present a derivation of the Heun solutions (and its confluences) from first principles  with no reference to supersymmetric gauge theory or localization, which sit in the background as inspiration to some of our ans\"atze, that nonetheless could be also accepted as such.
     Finally, we remark, that even if the main motivation for resuming PM and PN series relies on the study of open orbits, our hypergeometric representations provide an algorithm which is very efficient when implemented in a Mathematica code to describe PN expansions of bounded orbits. Indeed, here we compute the gravitational waveform and the energy flux, in the case of circular orbits, up to the 30th PN order\footnote{The ancillary mathematica files coming with this paper should help the reader to check our computations.}. This result will be tested against  the 22nd PN order computation available in literature \cite{Fujita:2012cm},  and results based on MST and numerical methods from the  Black Hole Perturbation Toolkit (BHTK) \cite{BHPToolkit}.  The analytic expressions of all ingredients contributing to the 30PN expansion of the wave form and energy flux are collected in ancillary mathematical files.  Numerical studies of unbounded systems are  postponed  to future investigations.

This is the plan of the paper: in Section \ref{sec2} we derive the PM and PN expansion of the confluent Heun function. In Section \ref{sec3} we apply the results of Section \ref{sec2} to the study of the gravitational waveforms emitted by a particle moving in a Schwarzschild background. In Section \ref{sec4} we test our results against those obtained via the MST method and numerical integration. In Appendix \ref{sec5} we derive the PM and PN expansions of the non-confluent Heun function.  Section \ref{sec6} contains our conclusions.

\subsection*{Conventions}

By confluent Heun equation we will refer to the ordinary differential equation (\ref{CHE}) of second order, with two regular singularities at $z=1,\infty$ and an irregular one at $z=0$. The equation is specified by five complex parameters dubbed here as $\{ m_1,m_2,m_3, u, x\}$.  The expansion of the general solution of this equation in the ''soft limit''  $x\ll 1$ is derived in section 2. These results apply to any physical problem governed by an equation of confluent Heun type. The focus of this paper is on the study of gravitational wave forms in
 Schwarzschild geometry. The Heun gravity dictionary in this case reads
   \beaq
x&=& 4{\rm i} \omega M \qquad , \qquad y=2{\rm i} \omega r  \qquad , \qquad z={2M\over r} \nn\\
m_1 &=&  \ft{x}{2} \quad ,\quad  m_2=m_3^*=-2+  \ft{x}{2}   \quad , \quad  u=(\ell+\ft12)^2+\ft{5x}{2} +\ft{x^2}{4}
\eeaq
 with $y=x/z$,  $M$ the mass of the BH,  $\omega$ and $\ell$ the frequency and orbital number of the gravitational wave  and $r$ the radial distance.
 Along the paper we set $G=1$ and use the mostly plus signature $(-+++)$.

 \section{The confluent Heun equation}\label{sec2}

The confluent Heun differential equation
\be
{\rm Hc}''(t)+ \left(
   \frac{\beta_0}{t}+\frac{\beta_1
   }{t-1} +\lambda\right){\rm Hc}'(t)+\frac{ \alpha_1 t -\alpha_0}{t(t-1) } {\rm Hc}(t)  =0 \label{CHE0}
\ee
 is an ordinary differential equation of second order with two regular singularities at $t=0,1$ and an irregular one at infinity. A basis of solutions is given by
 \beaq
Hc_+(t)&=&  {\rm HeunC} [\alpha_0,\alpha_1,\beta_0,\beta_1 ,\lambda, t] \\
Hc_-(t)&=&  t^{1-\beta_0 } \text{HeunC}[ \alpha_0+(1-\beta_0)(\lambda-\beta_1),\alpha_1 +\lambda(1-\beta_0 ) ,2-\beta_0 ,\beta_1 ,\lambda ,t] \nn
\eeaq
  with ${\rm HeunC}$ the confluent Heun function.

  \subsection{Confluent Heun-Gauge dictionary}

   The confluent Heun equation can be viewed as the quantum version of the Seiberg-Witten (SW) curve governing the dynamics of a ${\cal N}=2$ supersymmetric $SU(2)$ gauge theory with three flavours of masses $m_i$,  transforming in the fundamental representation of the gauge group\cite{Bianchi:2021mft,Bianchi:2021xpr} living on a curved Nekrasov-Shatashvili background \cite{Nekrasov:2009rc}.   The quantum curve can be written in the form\footnote{Equation (\ref{CHE}) can be written in the normal Schr\"odinger like form
   \be
\Psi''(z) + Q_{\rm gauge}(z)\Psi =0 \qquad , \qquad G_1(z)=e^{{\frac{x}{2 z}}} z^{m_3-1}(1-z)^{m_1+m_2-1\over 2} \Psi(z) \label{normal}
\ee
 with  	
 \be
 Q_{\rm gauge} = {-}\frac{x^2}{4 z^4}{+}\frac{x m_3}{z^3}{+}\frac{1{-} \left(m_1{-}m_2\right)^2}{4(z{-}1)z}+\frac{ 1{-}
 		\left(m_1{+}m_2\right){}^2}{4(z-1)^2 z}  +\frac{u{-}\ft{1}{4}{+}\ft{x}{2}
 		\left(m_1{+}m_2{-}1\right)}{(z{-}1)
 		z^2} \label{qgauge}
\ee		
   }
   {\small
   \be
\label{CHE}
 G_1''(z){+}G_1'(z) \left(\frac{1{-}m_1{-}m_2}{z{-}1}{-}\frac{2
   \left(m_3{-}1\right)}{z}{+}\frac{x}{z^2}\right)  {+}G_1(z)
   \left(\frac{\ft14{-} u{+}m_3(m_3{-}1)}{
   z^2}{+}\frac{u {+} \frac{3}{4} {+}\delta}{(z{-}1) z}\right)=0
   \ee
   }
   with $\delta = m_1 m_2 {+} m_1 m_3 {+} m_2 m_3 {-} m_1 {-} m_2 {-} m_3$, $x$ the gauge coupling and $u$ the  Coulomb branch parameter. (\ref{CHE}) is related to the standard form (\ref{CHE0}) of the Heun equation  by the Heun gauge dictionary
  \beaq
  G_1(z)  &=& z^{-{\alpha_1\over \lambda}}  {\rm Hc} ( \ft{1}{z}) \qquad , \qquad   t = z^{-1} \qquad, \qquad \lambda = -x \nn\\
   \alpha _0 &=& u-\ft14 +m_1(1-m_1) +(m_1+m_3-1) x  \qquad , \qquad \alpha_1= \left(m_1+m_3-1\right) x \nn\\
     \beta _0&=& 1 -m_1+m_2 \qquad , \qquad \beta_1=1 -m_1-m_2\label{CHE1}
  \eeaq
    Equivalently, (\ref{CHE}) can be written as a differential equation in the  variable $y=x/z$, taking the form
  {\small
   \be
\label{eq:CHE_y}
 G_0''(y){+}G_0'(y) \left(\frac{1{-}m_1{-}m_2}{y{-}x}{+}\frac{ m_1 {+} m_2 {+} 2m_3 {-} 1 }{y} {-}1 \right)   {+} G_0(y) \left(\frac{u{+}\ft34 {+} \delta}{(x-y) y}{+}\frac{1 {+} \delta {+} m_3(m_3{-}1)}{y^2}\right)=0
   \ee
   }
    with
\be
 G_1(\ft{x}{y})= g(x) G_0(y)
\ee
where $g(x)$ is a $y$-independent function. In these variables, the confluent Heun equations (CHEs)  (\ref{CHE}) and (\ref{eq:CHE_y})  have irregular singularities at  $z=0$ or $y=0$ and their solutions transform non-trivially  under rotations  around these points.  We denote by $G^0_\pm(y) $, $G^1_\pm(z) $, the eigenvectors of this action, i.e. solutions of the CHEs satisfying
  \be
G^0_\alpha(y\, e^{2\pi {\rm i}} )= e^{-2\pi {\rm i} \alpha a} G^0_\alpha(y) \qquad , \qquad  G^1_\alpha(z\, e^{2\pi {\rm i}} )= e^{2\pi {\rm i} \alpha a} G^1_\alpha(z)
\ee
 with $\alpha=\pm$ and $a(u,x,m_i)$ is an unknown function of the Heun parameters. The quantity $a$ will play a fundamental role in the following.  One of the lessons of the Heun gravity correspondence is that the solutions of the CHE and the connection matrices look simpler if expressed in terms of $a$ rather than in terms of $u$.

     We will focus on the "soft limit" where $x$ is taken small (large wave lengths), keeping finite either the variable $z$ or $y$.  We will refer to these two limits as PM and PN respectively.
   The two choices  lead to equations of hypergeometric type. Indeed, sending  $x\to 0$ in  \eqref{CHE} and (\ref{eq:CHE_y}) lead to hypergeometric equations with solutions
 \beaq
{\rm PM} : && \qquad G^0_\alpha(y)   \underset{x\to 0\atop y~{\rm finite}}{\approx}  H^0_\alpha(y)  \nn\\
{\rm PN} : && \qquad G^1_\alpha(z) \underset{x\to 0\atop z~{\rm finite}}{\approx}  H^1_\alpha(z)   \label{lim12}
 \eeaq
   where
\beaq \label{h012}
H^0_\alpha(y) &=& y^{\frac{1}{2}-\alpha a-m_3} \,
   _1F_1\left(\ft{1}{2}-\alpha a-m_3;1-2\alpha
   a;y\right) \nn\\
H^1_\alpha(z) &=& z^{-\frac{1}{2}+\alpha a+m_3} \,
   _2F_1\left(\frac{1}{2}+\alpha a-m_1,\frac{1}{2}+\alpha a-m_2;1+2\alpha
   a;z\right)
\eeaq
  At this order,   $a \underset{x\to 0}{\approx} \sqrt{u}$.  For $ x\neq 0$ but small one finds
 \beaq
{\rm PM}_k : && \qquad G^0_\alpha(y)  = y^{\frac{1}{2}-\alpha a-m_3 } \sum_{i=0}^k \sum_{j=0}^\infty  c^{PM}_{ij}( \alpha a) x^i  y^{j-i}  +O(x^{k+1}) \nn  \label{defsigma}\\
{\rm PN}_k : && \qquad G^1_\alpha(z)  = z^{-\frac{1}{2}+\alpha a+m_3 }\sum_{j=0}^k\sum_{i=0}^\infty c^{PN}_{ij}( \alpha a) z^{i-j} x^{j}  +O(x^{k+1})
 \eeaq
and
\be
u = a^2+\sum_{i=1}^k u_i(a) x^i
\label{uvsa}\ee
 with the coefficients $c^{PM}_{ij}(  a)$, $c^{PN}_{ij}( a)$, $u_i(a)$   determined  by solving the CHE equations  order by order in $x$. Inverting (\ref{uvsa}) one finds the dependence of the monodromy eigenvalue $a(u)$ on the parameter $u$. Moreover,  since $G^1_\alpha(z)$ and $ G^0_\alpha(\ft{x}{z})$ are solutions of the same equation and have the same monodromy around $z=0$ they should be proportional to each other.
 We conclude therefore that in the overlap region, the so called  {\rm near zone}, where $x\ll z \ll 1$ we find
\be
{\rm Near~zone}: \qquad  G^1_\alpha(z) = g_{\alpha}(x) G^0_\alpha(\ft{x}{z})
\ee
The main result of this Section is that,  the  series PM$_k$  and PN$_k$  in (\ref{defsigma}) can be exactly resummed and written in terms of a single hypergeometric function. This will  provide us with a first principle proof of the connection formulae for the confluent Heun functions, where the whole non-triviality of the Heun equation is codified into a single function $a(u)$.

  For concreteness, we will focus on the domain $z\in [ 0, 1]$. The general  solution $G_1(z)=\sum_\alpha c_\alpha G^1_\alpha(z)$  behaves   near the boundaries as
 \beaq
  G_1(z) &\underset{z\to 0}{\approx}&  B_{-} (1+\ldots) +B_{+}   z^{ 2 m_3}  e^{ x\over z}  (1+\ldots) \nn\\
  G_1(z) &\underset{z\to 1}{\approx}&  D_{-}(1+\ldots)  +D_{+} (1-z)^{m_1+m_2} (1+\ldots) \, \, \, \, \, \, \, \, \,  \label{gbcgen0}
\eeaq
  where the dots stand for subleading terms in these limits. Specifying boundary conditions corresponds to fix two linear combinations made out of the four coefficients $B_{\pm}$,  $D_{\pm}$. We will derive connection formulae   relating the coefficients $B_\pm$ to  $D_\pm$ and viceversa.

\subsection{PM approximation: exterior region}

In this subsection we consider the limit $x\ll 1$ keeping $y$ finite. This corresponds to the PM approximation and we refer to this region as \textit{the exterior region}: the reason of it will be clear in the next Section from the dictionary with the gravity side (see \eqref{rdict}).\\
We denote the basic solution as $G_\alpha^0(\ft{x}{z})$ and at leading order in $x$ we take
  \be
  G^0_\alpha(y)=H^0_\alpha(y)
 \ee
   For higher orders in $x$, proceeding by trial and error, we come to the following ans\"atze
\be
    G^0_\alpha(y) =     P_{0}(y) \, H^0_\alpha(y) {+} \widehat{P}_{0}(y) \,  y H^{0}_\alpha{}'(y)   \label{g00}
\ee
where the prime stands for derivation and
 \beaq
P_{0 }( y)  &=& 1+\sum_{i=1}^k   \sum_{j=0}^{i-1}   c_{ij} \, x^i y^{j-i}  \nn\\  \widehat{P}_{0 }(y)   &=& \sum_{i=1}^k   \sum_{j=0}^{i-1}   \widehat{c}_{ij} \, x^i y^{j-i}   \label{pp0bis}
\eeaq
are polynomials of order $k$ in $x$\footnote{For $k=1$ M.Billò, M.L.Frau and A.Lerda first obtained \eqref{g00} resumming the contributions of the Young tableaux entering the partition function of the supersymmetric gauge theory.}. We plug this ans\"atze into (\ref{CHE}) and use the differential equation to express higher derivatives of $H^0_\alpha$ in terms of $H^0_\alpha$ and its first derivative themselves. Therefore one finds a system of algebraic linear equations that determines the coefficients $c_{ij}$, $\widehat{c}_{ij}$  and $u_i$ (whose complete forms are given in \eqref{csol}).
Setting $c_{00}=1$, one finds, for the first term in the recursion
\beaq
 c_{10} &=& u_1=\frac{1}{2}(1-m_1-m_2-m_3)  -{2 m_1 m_2 m_3\over 4a^2-1} \nn\\
 \widehat{c}_{10}& =&\partial_{m_3} u_1=-\frac{1}{2}   -{2 m_1 m_2 \over 4a^2-1}   \label{cus}
 \eeaq
  It is easy to check that $c_{ij}$, $\widehat{c}_{ij}$ are invariant under $a\to -a$. This is true at any PM order, so all the dependence on $\alpha$ of the solution is codified in $H^0_\alpha$.
 Explicit computations show that the coefficients $c_{i,i-1}$, $\widehat{c}_{i,i-1}$  can be always written in terms of derivatives of $u_i$. Indeed, one finds
  \beaq
  & c_{i,i-1} = u_i  \, , \nn\\
  & \gamma_m = 1+\sum_{i=1}^\infty  \widehat{c}_{i,i-1}x^i = e^{-\partial_{m_3}  {\cal F}_{\rm inst} } \quad \label{dk1f}
  \eeaq
 with
  \be
   {\cal F}_{\rm inst}  \equiv  -\sum_{i=1}^\infty {u_i \over i} x^i \label{finst}
  \ee
  The subscript "inst" reminds the reader that this quantity is interpreted as the instanton prepotential in the gauge theory description of the Heun equation.
  Indeed in the limit $z\to 0$, where one of the two gauge couplings is turned off, the quiver theory reduces to a  ${\cal N}=2$ supersymmetric SU(2) gauge theory with three massive  fundamental flavours with prepotential ${\cal F}_{\rm inst}$.

We notice that  (\ref{g00}) is exact in $y$  at the k-th PM order, so we can extrapolate it to the region of large $y$ using the standard hypergeometric transformation rule
 \beaq
H^0_\alpha (y) = \sum_{\alpha'=\pm}  B_{\alpha \alpha'}  \widetilde{H}^{0}_{\alpha'}(y)
\label{braiding0}\eeaq
 with
 \be
   \widetilde{H}^0_{\alpha}(y) =y^{-m_3(1+\alpha ) }
e^{ y (1+\alpha)\over 2} {}_2 F_0 \left(  \frac{1}{2}-a+\alpha  m_3,\frac{1}{2}+a+\alpha
   m_3,\frac{\alpha }{y}\right)
   \ee
and
  \be
  B_{\alpha \alpha'} =e^{\frac{{\rm i} \pi}{2}  \left(1-\alpha' \right) \left({1\over 2}  -\alpha a - m_3 \right)}   \frac{\Gamma (1-2 \alpha a  ) }{\Gamma
   \left(\frac{1}{2}-\alpha a  -\alpha' m_3 \right)}
  \ee
  We can therefore introduce an equivalent basis of solutions, naturally defined in the \textit{far region}.
  \be
  \widetilde{G}^0_{\alpha}(y) =   P_{0}(y)   \widetilde{H}^{0}_{\alpha}(y)   {+} \widehat{P}_{0}(y)   y \widetilde{H}^{0}_{\alpha}{}'(y)
\ee
 related to the previous one by the connection formulae
\be\label{conn_form_B}
 G^0_\alpha(y) =   \sum_{\alpha'=\pm}  B_{\alpha \alpha'}  \widetilde{G}^{0}_{\alpha'} (y)
\ee

\subsection{PN approximation: interior region}

In this subsection we consider the limit $x\ll 1$ keeping $z$ finite. This corresponds to the PN approximation and we refer to this region as \textit{the interior region}. The analysis proceeds {\it mutatis mutandis} following the same steps of the previous subsection, exchanging the roles of $y$ and $z$.\\
We denote by $G^1_\alpha(z)$ the basic solutions and at leading order in $x$ we  take
  \be
  G^1_\alpha(z)=H^1_\alpha(z)
 \ee
   Higher orders  follow from the ans\"atze
\be
    G^1_\alpha(z) =     P_{1}(z) \, H^1_\alpha(z) {+} \widehat{P}_{1}(z) \,  z H^{1}_\alpha{}'(z)   \label{g0}
\ee
where the prime stands for derivation and
 \beaq
P_{1 }(z) & = &1+\sum_{j=1}^k   \sum_{i=0}^{j-1}   d_{ij} \, z^{i-j} x^j  \nn\\ \widehat{P}_{1 }(z)   &=& \sum_{j=1}^k   \sum_{i=0}^{j}   \widehat{d}_{ij} \, z^{i-j} x^j \label{pp1bis}
\eeaq
Setting $d_{00}=1$, one finds that the first non-trivial coefficients are
\beaq
 d_{01} =\frac{1}{2}+ \frac{2 m_3\left(1-m_3\right) }{4 a^2-1} , \quad
 \widehat{d}_{01} =- \widehat{d}_{11}=\frac{  2 m_3}{4 a^2-1}
  \eeaq
  We notice that the results are invariant under $a\to -a$. This is true at any PN order, so all the dependence on $\alpha$ of the solution is codified in $H^1_\alpha$.

  Proceeding as before, we can extrapolate the PN representation to the region $z\approx 1$ using the  hypergeometric transformation rule
   \beaq
  H^1_\alpha(z) = \sum_{\alpha'=\pm}  F_{\alpha\alpha' } \widetilde{H}^1_{\alpha'}(z)
\label{fusion0}\eeaq
with
  \be
\widetilde{H}^{1}_{\alpha}(z)  = z^{a{+}m_3{-}\frac{1}{2}} (1{-}z)^{\frac{1}{2} \left(m_1{+}m_2\right)
   \left(1{+}\alpha \right)} \, _2F_1\left(\frac{1}{2}{+}a{+}\alpha m_1, \frac{1}{2}{+}a{+}\alpha m_2 ;1{+}\alpha\left(m_1{+}m_2\right) ;1{-}z\right)
\ee
and
\be
   F_{\alpha\alpha'}  = \frac{\Gamma (1{+}2\alpha a ) \Gamma
   \left(-\alpha' \left(m_1+m_2\right) \right)}{\Gamma \left( \ft12+\alpha a  -\alpha' m_1  \right) \Gamma \left( \ft12 +\alpha a  -\alpha' m_2  \right)}
\ee
An equivalent  basis of solutions defined in the \textit{near horizon region} is therefore given by
\be
   \widetilde{G}^1_{\alpha}(z) =   P_{1}(z) \, \widetilde{H}^{ 1}_{\alpha}(z)   {+} \widehat{P}_{1}(z) \,  z\, \widetilde{H}^{ 1 }_{\alpha}{}'(z)
\ee
 satisfying the connection formulae
\be\label{conn_form_F}
   G^1_\alpha(z) = \sum_{\alpha'=\pm}  F_{\alpha\alpha' } \widetilde{G}^1_{\alpha'}(z)
\ee

 \subsection{Boundary conditions and connection formulae}

  The functions $G^0_\alpha(\ft{x}{z})$ and $G^1_\alpha(z)$ are solutions of the same differential equation and have the  same monodromy around $z=0$, so they should be proportional to each other. We denote by $g_\alpha(x)$ their ratio
 \begin{equation}
  g_{\alpha}(x)={ G^1_\alpha(z)\over G^0_\alpha (\ft{x}{z}) } =x^{\alpha a{+}m_3{-}\frac{1}{2}} \left[ 1{+}x \left(\ft{1}{4}{-}\ft{m_3 \left(m_1{+}m_2{+}2
   m_3{-}2\right){-}m_1 m_2}{4 a^2-1} {+} \ft{8 \, \alpha \, a \, m_1 m_2
   m_3}{\left(4 a^2{-}1\right)^2} \right)+\ldots  \right ] \label{galpha}
\end{equation}
Expanding for small $z$ and $y$ one can check explicitly that this ratio depends only on $x=z y$. Furthermore one finds
\be
 \gamma(x) ={g_+(x)\over g_-(x) }=   x^{2a} e^{-\partial_a {\cal F}_{\rm inst}(a,x) }   \label{gamma2}
\ee
with ${\cal F}_{\rm inst}(a,x)$ defined in (\ref{finst}). Collecting all the different pieces together, we have at our disposal four different basis of solutions $\{ G^0_\alpha, \widetilde{G}^0_{\alpha}, G^1_\alpha, \widetilde{G}^1_{\alpha} \}$
related to each other by the connection matrices $B_{\alpha \alpha'}$, $F_{\alpha \alpha'}$. More precisely, the general solution can be written as
\beaq
G_1(z) &=&  \sum_\alpha c_\alpha g_\alpha(x)  G^0_\alpha(\ft{x}{z} )    = \sum_{\alpha,\alpha'}  c_\alpha g_\alpha(x) B_{\alpha \alpha'}  \widetilde{G}^0_{\alpha'}(\ft{x}{z} ) \nn\\
&=&  \sum_\alpha c_\alpha   G^1_\alpha(z )   = \sum_{\alpha,\alpha'}  c_\alpha  F_{\alpha \alpha'}  \widetilde{G}^1_{\alpha'}(z )   \label{g4}
\eeaq
The first line corresponds to the PM approximation where $x$ is taken small keeping $y$ finite. The second equality in this line is obtained using the connection formulae  \eqref{conn_form_B}.
Similarly, the second line corresponds to the PN approximation where  $x$ is taken small keeping $z$ finite and the right hand side follows from  \eqref{conn_form_F}.
  We stress that these connection formulae are exact. As in the case of the hypergeometric connection formulae , they can be written as ratios of gamma functions with
   the {\it caveat} that the arguments of these functions depend on $a(u)$. The whole non-triviality of the confluent Heun equation is codified in a single function  $a(u)$.
  Expanding at the leading order the two right hand sides in \eqref{g4} and using (\ref{dk1f}), one finds the asymptotics
  \beaq
 G_1(z ) & \underset{z\to 0}{\approx} & \sum_{\alpha,\alpha'}  c_\alpha g_\alpha(x) B_{\alpha \alpha'}   (\ft{x}{z})^{ -m_3 (1+\alpha')    }  e^{ x (1+\alpha')\over 2 z}  e^{ - \frac{ 1+\alpha'}{2} \partial_{m_3} {\cal F}_{\rm inst}   } \nn \\
 G_1(z)& \underset{z\to 1}{\approx}  & \sum_{\alpha,\alpha'}  c_\alpha  F_{\alpha \alpha'}   (1{-}z)^{{1{+}\alpha'\over 2 }(m_1{+}m_2)} \, h_{\alpha'}\label{ginf}
 \eeaq
   with
    \be
  h_{\alpha'}=    P_{1}(1) \,    {+} {1{+}\alpha'\over 2 }(m_1{+}m_2)   \widehat{P}_{1}{}'(1) \,
   \ee
   Once again the ratio of these coefficients takes a simple form
   \be
  \frac{h_+}{h_-}=e^{\partial_{m_1+m_2}{\cal F}_{\rm inst}}
  \label{hh} \ee
  Comparing against (\ref{gbcgen0}) one finds the  asymptotic coefficients
 \beaq
 B_{\alpha'} & =&   \sum_{\alpha}  c_\alpha g_\alpha(x) B_{\alpha \alpha'}  x^{ -m_3 (1+\alpha' )  }    e^{- \frac{ 1+\alpha'}{2} \partial_{m_3} {\cal F}_{\rm inst}    } \nn\\
 D_{\alpha'} & =&   \sum_\alpha c_\alpha  F_{\alpha \alpha'}  h_{\alpha'}  \label{bdcoef}
 \eeaq

\section{Black hole perturbation}\label{sec3}

In this Section we apply the results obtained in the previous one to the study of gravitational waves in the Schwarzschild geometry.

\subsection{Heun gravity correspondence}

BHs perturbations are typically described by equations of the confluent Heun type. For example, the Teukolsky equation for spin $s$ perturbations in the Schwarzschild geometry is given by
\be
{1\over \Delta (r)^{s} }{d\over dr} \left[ \Delta (r)^{s+1} R'(r) \right] { +}\left(\frac{\omega^2 r^4{-}2{\rm i} s (r{-}M)  \omega r^2  }{\Delta (r)}{+}4{\rm i} s \omega r{-} (\ell{-}s)(\ell{+}s{+}1) \right)R(r)  = 0 \label{teuk}
\ee
with $\Delta=r(r-2M)$.
We are interested in the $\psi_4$ fundamental mode \cite{Newman:1961qr,Teukolsky:1973ha} corresponding to $s=-2$. (\ref{teuk}) can be put into the CHE form (\ref{CHE}) after the following identifications
\beaq
R (z) &=&  e^{-{x\over 2z}} (1-z)^{ 2 {-}  {x\over 2}    }  (\ft{x}{z})^{ {-}1{-}{x\over 2}}   G_1(z)  \qquad , \quad z= {2M\over r}  \qquad, \qquad x=4 i M \omega \label{rdict} \\
m_1 &=&  \ft{x}{2} \quad ,\quad  m_2=m_3^*=-2+  \ft{x}{2}   \quad , \quad  u=(\ell+\ft12)^2+\ft{5x}{2} +\ft{x^2}{4}\nn
\eeaq
We denote by  $R_{\rm in} (z)$ a solution satisfying incoming boundary conditions at the horizon and by $R_{\rm up} (z) $ that one satisfying upgoing boundary conditions at infinity, i.e.
\beaq
R_{\rm in} (z)  & \approx &
\left\{
\begin{array}{lll}
B^{\rm in}_{-}   e^{-{x\over 2z}}  z^{-1-m_3 }    +  B^{\rm in}_{+}   e^{ {x\over 2z}}  z^{ m_3-1} &~~~~~~~   &\quad z \to 0   \\
D^{\rm in}_- 	 (1-z)^{  1-{m_1+m_2\over 2}   }    &   & \quad
	z\to 1    \\
\end{array}
\right.  \nn\\
R_{\rm up} (z)  & \approx &
\left\{
\begin{array}{lll}
   B^{\rm up}_{+}   e^{ {x\over 2z}}  z^{ m_3-1} &~~~~~~~   &\quad z \to 0   \\
D^{\rm up}_- 	 (1-z)^{  1-{m_1+m_2\over 2}   }  +D^{\rm up}_+ (1-z)^{  1+{m_1+m_2\over 2}   }  &   & \quad
	z\to 1    \\
\end{array}
\right.  \label{bcpsi}
\eeaq
where we have set to zero the coefficients  $D^{\rm in}_+$ and $B^{\rm up}_-$.
In radial coordinates $R_{\rm in,up} (r)\equiv  R_{\rm in,up} \left[ z(r)\right] $.
In terms of these solutions the Green function reads
\be
   G(r,r')={1\over W} \left\{
\begin{array}{ccc}
R_{\rm in} (r')   R_{\rm up} (r)  & ~~~~~~~&  r'< r \\
R_{\rm in} (r)   R_{\rm up} (r')  & ~~~~~~~ &   r< r' \\
\end{array}
\right.\label{green}
\ee
with
\be
W={R_{\rm in} (r)   R'_{\rm up} (r)- R'_{\rm in} (r)   R_{\rm up} (r) \over \Delta(r) } ={ {\rm i} \omega \over 2 M^2}  B^{\rm in}_- B^{\rm up}_+
\label{ww}
\ee
a constant built out of the Wronskian which is conveniently computed for $r\to \infty$,
since its first derivative is zero as it follows using (\ref{teuk}).
In the following we use as independent variables the two dimensionless combinations
\beaq
  y & =& 2  {\rm i}  \omega r  \qquad , \qquad    x= 4{\rm i} \omega M  \label{ekappa}
\eeaq
The incoming and upgoing solutions can be written as linear combinations
\be
  R_{\rm in,up}(y)    =     \sum_{\alpha} c^{\rm in,up}_\alpha g_\alpha  R_{\alpha}(y)
   \ee
 of the two independent solutions of the Teukolsky equation, let us say in the PM representation, that according to (\ref{rdict}) are given by
\be
 R_\alpha(y) =e^{- {y\over 2  } } \left(1- \frac{x}{y}\right)^{ 2 {-}  {x\over 2}    }  y^{ {-}1{-}{x\over 2}}   G^0_{\alpha}(y)
   \label{rka}
\ee
    $c^{\rm in,up}_\alpha$ are some coefficients, chosen such that the incoming boundary conditions  at the horizon or upgoing  at infinity are satisfied,
i.e.
\be
 D^{\rm in}_{+} =B^{\rm up}_{-} =0
\ee
 with asymptotic coefficients given by (\ref{gbcgen0}), (\ref{rdict})
 \beaq
 B^{\rm in,up}_{\alpha'} & =&   \sum_{\alpha}  c^{\rm in,up}_\alpha g_\alpha(x) B_{\alpha \alpha'}  x^{ 1- \alpha' m_3     }    e^{ -\frac{ 1+\alpha'}{2} \partial_{m_3} {\cal F}_{\rm inst}    }
       \nn\\
 D^{\rm in,up}_{\alpha'} & =&   \sum_\alpha c^{\rm in,up}_\alpha  F_{\alpha \alpha'}  h_{\alpha'}  e^{- {x\over 2} }  \,   x^{1+m_3}  \label{bdcoef2}
 \eeaq
We are interested in the PN expansion of the solutions, since we want to describe binary systems in the limit where the distance between the two objects is large and velocities small. For example, for circular orbits with tangential velocity $v$, this corresponds to the limit where $v$ is small,  $y \sim v$ and $x\sim v^3$.
Therefore the PN expansion can be written as
\be \label{PNexpRa}
 R_\alpha (y) = y^{{3\over 2}-\alpha a} \sum_{n=0}^\infty  R_n( \alpha a, y)
\ee
with
\beaq
 &&  R_0(a,y)= 1 \nn\\
&& R_1(a,y)=\frac{2 y }{1-2 a}\nn\\
&& R_2(a,y)=\frac{(17-2 a) y ^2}{ 16(2 a-1)(a-1)}+\frac{(2 a-3) x}{4 y }\nn\\
 && R_3(a,y)=
  \frac{(7-2 a) y ^3}{8(2a-1)(2a-3)(a-1)}+\frac{(11-6 a) x}{4(2
   a-1)} \label{rfinaln} \nn\\
  &&R_4(a,y)=  \frac{\left(4 a^2{-}72 a{+}163\right) y ^4}{512 (a{-}2) (a{-}1) (2 a{-}3) (2
   a{-}1)}\nn\\
   &&+\frac{\left({-}4 a^2{+}48 a{-}119\right) y  x}{64 (2 a{-}1)(a{-}1)}{+}\frac{\left(8 a^3{-}4 a^2{-}18 a{+}9\right) x^2}{ 64y ^2 (a{+}1) }
\label{Rexp}\eeaq
and
\be
   a(\ell)=\ell+\frac{1}{2}  +\frac{\left(15 \ell^4+30 \ell^3+28 \ell^2+13 \ell+24\right)  x^2}{8 \ell (\ell+1) (2 \ell+1) \left(4 \ell^2+4 \ell-3\right)}+\ldots \label{als}
\ee

\subsection{The incoming solution}

 The incoming solution is obtained by requiring no outgoing waves at the horizon, i.e. by setting $D^{\rm in}_+=0$. According to (\ref{bdcoef2}), such condition leads to
\be
  \sum_{\alpha=\pm}  c^{\rm in}_\alpha F_{\alpha +}   =0 \label{dp0}
\ee
This determines the ratio  of the two coefficients to be
\be
\ {c^{\rm in}_+\over  c^{\rm in}_- } =- { F_{-+}   \over F_{++}   } =
\frac{\Gamma({-} 2 a) \Gamma \left(  \ft12{-} m_1    {+}a \right)
\Gamma \left(   \ft12{-} m_2 {+}a \right) }{\Gamma(2 a) \Gamma \left(   \ft12{-}m_1{-}a  \right)
\Gamma \left( \ft12{-} m_2 {-} a  \right)}   \label{cinup}
\ee
On the other hand, using the first of (\ref{ginf}), one finds the asymptotic behaviour at infinity
\be
  R_{\rm in}(y)    \underset{y \to \infty} {\approx}       \sum_{\alpha,\alpha'} c^{\rm in}_\alpha g_\alpha  B_{\alpha \alpha'}
   e^{ \alpha' y \over 2 }   y^{1-\alpha'  m_3   }  e^{  -\frac{ 1+\alpha'}{2} \partial_{m_3} {\cal F}_{\rm inst}   }
     \label{rka2}
 \ee
Comparing against the first line of $R_{\text{in}}$ in (\ref{bcpsi}), one finds
\be
 B^{\rm in}_{\alpha'}  =    \sum_{\alpha} c^{\rm in}_\alpha g_\alpha  B_{\alpha \alpha'}
     x^{1 -\alpha' m_3  }  e^{  -\frac{ 1+\alpha'}{2} \partial_{m_3} {\cal F}_{\rm inst}   }    \label{binalphap}
\ee
We notice that the  incoming boundary condition determines only the ratio  $c^{\rm in}_+/c^{\rm in}_-$, leaving undetermined the overall normalization $c^{\rm in}_-$. This dependence however cancels out in the Green function after dividing by the factor $W$ \eqref{ww}. It is convenient therefore to define a ratio which does not depend on overall normalizations
\be
 \mathfrak{R}_{\rm in} (y)=  { R_{\rm in} (y)  \over  x  B^{\rm in}_-}
 =   C_{\rm in}   \left[  R_-(y)  - \gamma  \ft{F_{-+}}{F_{++}}  R_+(y)  \right]  \label{rgg}
\ee
with
\be
C_{\rm in}  = {    x^{x\over 2 } \over B_{--} }
\left( 1-  \gamma    \ft{F_{-+}  B_{+-} }{ F_{++}B_{--}   } \right)^{-1} \underset{x\to 0}{\approx}(-1)^{ \ell+1  }
 \frac{\Gamma \left(  \ell{-} 1  \right)}{   \Gamma (2  \ell{+}2 )   }  +\ldots
\label{Cin}\ee
We notice that in the PN limit $ R_\alpha (y) \sim  y^{{3\over 2}-\alpha a}$ and $\gamma \sim x^{2a}$, so the $\gamma R_+$-contribution in (\ref{rgg}) is suppressed by an extra factor $(x/y)^{2a}$ with respect to  $R_-$. The incoming solution is therefore dominated by the $R_-$ component. Using (\ref{als}) one finds the PN expansion
{
	\beaq
	&&   R_-(y)=y^{a+{3\over 2} } \left[  1{+}\frac{ y }{1{+}\ell} {+} \left( \frac{(9{+}\ell) y
		^2}{8 (1{+}\ell) (3{+}2\ell)} {-} \frac{(\ell{+}2) x}{2 y }\right){+} \left(\frac{(\ell{+}4) y ^3}{8 (\ell{+}1) (\ell{+}2) (2 \ell{+}3)} {-}\frac{(7{+}3 \ell) x}{4
		(\ell+1)}\right)  \right. \nn\\
	&& \left. +
	\left(\frac{\left(50{+}19 \ell{+}\ell^2\right) y
		^4}{128 (\ell{+}1) (\ell{+}2) (2 \ell{+}3) (2 \ell{+}5)} {+} \frac{(\ell^2-1)   ( \ell+2) x^2}{4 (2\ell-1) y ^2}-\frac{(\ell+4)
		(\ell+9) x y }{16 (\ell+1) (2 \ell+3)}\right) +\ldots \right]
	\eeaq
}
It is important to observe that at higher orders both $R_\pm(y)$ exhibit  poles in the limit where $\ell \to \mathbb{Z}$, but one can check that they cancel against each other in the combination (\ref{rgg}).
For example, for $\ell=2$ the poles  in the two components first appear at order $v^{16}$.

\subsection{The upgoing solution}

Upgoing boundary conditions correspond to setting $B^{\rm up}_-=0$,  that according to (\ref{bdcoef2}) leads to
\be
{c_-^{\rm up} \over c_+^{\rm up}}  = -\gamma {B_{+-}\over B_{--}}
\ee
Following the same passages of the incoming solution, we can write the asymptotic behaviour of $R_{\text{up}}(y)$ at infinity, which is the same as \eqref{rka2} but with $c^{\text{up}}_\alpha$ in place of $c^{\text{in}}_\alpha$. This implies that also the coefficients  $B_\alpha^{\text{up}}$ take the same form of \eqref{binalphap} with the same substitution $c^{\text{in}}_\alpha \to c^{\text{up}}_\alpha$. Again we introduce the ratio
\be
   \mathfrak{R}_{\rm up} (y)=(2M)^3 {R_{\rm up} (y)  \over  B^{\rm up}_+} =    C_{\rm up}    \left[  R_+(y)   -  \ft{B_{+-}}{B_{--}}  R_-(y)  \right] \label{rupfinal}
\ee
with
\beaq
  C_{\rm up}&=&   { (2M)^3  e^{\partial_{m_3} {\cal F}_{\rm inst} } \over    x^{3+\frac{x}{2}} B_{++} }     \left( 1-\ft{B_{- +} B_{+-}}{B_{+ +}B_{- -}} \right)^{-1}  = \frac{(2M)^3}{x^{3+\frac{x}{2}}}   \frac{\Gamma(2-\ell)}{\Gamma(-2\ell)}   +\ldots
\eeaq
We notice that the leading term on the right hand side is finite in the limit where $\ell$ becomes an integer with $\ell \geq 2$. The same result is obtained by setting $\ell  \to \mathbb{N}$ from the very beginning and sending then $x \to 0$.
In the PN limit, the solution is dominated by the $R_+(y)$ component with PN expansion
\beaq
&& R_+(y)=  y^{{3\over 2}-a}  \left[  1{-}\frac{ y }{\ell}{+} \left(\frac{(\ell-1) x}{2 y }{-}\frac{(\ell{-}8) y
	^2}{8 l (2\ell{-}1)}\right){+} \left(-\frac{(3\ell{-}4) x}{4
	\ell}{+}\frac{(\ell{-}3) y^3}{8 (\ell{-}1) \ell (2\ell{-}1)}\right) \right. \nn\\
&& \left. +
\left(\frac{( \ell{-}1) \ell (\ell{+}2) x^2}{4 (2\ell{+}3) y ^2}-\frac{(\ell{-}8) (\ell{-}3)
	x y }{16 \ell (2\ell{-}1)}+\frac{\left(32{-}17 \ell{+}\ell^2\right) y ^4}{128
	(\ell{-}1) \ell (2\ell{-}3) (2\ell{-}1)}\right) \right]
\eeaq

  It is important to observe that again the poles in $R_\pm(y)$  in the limit where $\ell \to \mathbb{Z}$ cancel against each other in the combination (\ref{rupfinal}).
For $\ell=2$ the poles in the two components first appear at order $v^{7}$.

\section{Wave form and energy flux at 30PN}\label{sec4}

In this Section we use perturbation theort to compute the probe limit of the wave form and the energy flux produced by a circular EMRI binary system  at order 30PN, i.e. $ v^{60} $, with $v$ the tangential velocity of the light particle.
The results will be compared against the 22PN results available  in the literature \cite{Fujita:2012cm} and the numerical computations using the package \textit{Teukolsky} of the Black Hole Perturbation Toolkit (BHTK) \cite{BHPToolkit} based on the MST method \cite{Tagoshi:1993dm,Poisson:1994yf,Shibata:1994jx,Tagoshi:1994sm,Tanaka:1996lfd,Mano:1996vt,Mino:1997bx,Suzuki:1998vy,Fujita:2010xj,Fujita:2011zk} and on a numerical integrations of the differential equation.
We stress the fact that by  30PN, we mean that the polynomials coefficients $P_0$, $\widehat{P}_0$ and the exponent $a$, defining the solution are truncated at order 30PN, but the hypergeometric functions and the $c^{\rm in}_\alpha(a)$ coefficients specifying the solution are kept exact, so an infinite number of terms beyond 30PN are taken into account. In particular, the infinite tower of tail contributions (log divergences, $\pi$'s and transcendental numbers) are resummed into  exponentials and Gamma functions. The observables we consider are  the $(\ell,m)$ harmonic modes of the incoming solution $\mathfrak{R}_{\text{in}}$ and the energy flux radiated to infinity by the system.

Since for circular orbits $v^2=M/r$, and  $r$ should be taken bigger than the innermost stable circular orbit (ISCO) radius  $r_{\text{ISCO}} = 6 M$,  the domain of variability of the tangential velocity is $v\in[ 0,\ft{1}{\sqrt{6}}] $. We find that the first 12 digits of 30PN results are stable  for any choice $v$ in this domain, so we will always display this number of digits.  The numerical computations and those with the MST method are performed  setting in the BHTK Mathematica files: AccuracyGoal = 1000, PrecisionGoal = 70, WorkingPrecision = 80,  SetPrecision=80.  With this choice of parameters both numerical and MST results agree to a 12 digits precision. In our tables these results will be denoted as BHTK.

\subsection{The incoming and upgoing solutions}

In this subsection we collect all the ingredients needed in the computations of the incoming, upgoing waveforms at order k=30 in the hypergeometric PM representation.
We write\footnote{ Here we rewrite the second terms in the right hand sides of \eqref{rgg}, \eqref{rupfinal},  as the images under $a \to -a$ of the first terms,  using the transformation rules
    $R_- \to R_+$, $\gamma \to \gamma^{-1}$,  $F_{-\alpha'} \leftrightarrow  F_{+ \alpha'}$ and $B_{-\alpha'} \leftrightarrow  B_{+ \alpha'}$. }
\beaq
   \mathfrak{R}_{\rm in} (y) &=&      C_{\rm in}    R_-(y) + (a \to -a)  \nn\\
    \mathfrak{R}_{\rm up} (y) & =&     C_{\rm up}    R_+(y) + (a \to -a) \label{rinrup}
\eeaq
with\footnote{We  notice that  $\gamma_{\rm tot} = \gamma  \ft{F_{-+}  B_{+-} }{ F_{++}B_{--}}  = e^{-2\pi {\rm i} a_D}= e^{-\partial_a  ({\cal F}_{\rm tree}+{\cal F}_{\rm one-loop}+{\cal F}_{\rm inst}) } $ collects the  tree level, one-loop and instanton contributions to the dual SW period $a_D$.  }
\beaq \label{collection1}
R_\alpha (y) &=& e^{- {y\over 2  } } \left(1- \ft{x}{y}\right)^{ 2 {-}  {x\over 2}    }  y^{ {-}1{-}{x\over 2}}    \left[  P_{0}( y) \, H^0_\alpha(y) {+} \widehat{P}_{0}( y) \,  y H^{0}_\alpha{}'(y)   \right] \nn\\
H^0_\alpha(y) &=& y ^{\frac{5}{2} +\frac{x}{2}-\alpha a } \, _1F_1\left( \ft52 +\ft{x}{2}- \alpha a   ;1-2 \alpha a;y \right)\nn\\
C_{\rm in}  &=& x^{x\over 2 } e^{ -\frac{{\rm i} \pi}{2}(5+2a+x) }  {    \Gamma(-\ft32+a-\ft{x}{2} )\over \Gamma(1+2a) }
\left( 1- \gamma_{\rm tot}  \right)^{-1}  \nn\\
C_{\rm up}  &=&    { (2M)^3    \over    x^{3+\frac{x}{2}}   \, \gamma_m }   {    \Gamma( \ft52-a+\ft{x}{2} )\over \Gamma(1-2a) }   \left( 1-e^{-2\pi {\rm i} a} \, \ft{\cos \ft{\pi}{2}(x+2a)}{\cos \ft{\pi}{2}(x-2a)}  \right)^{-1}\nn\\
 \gamma_m  &=&1+\sum_{i=1}^\infty  \widehat{c}_{i,i-1} x^i   \quad , \quad  \gamma =x^{2a} e^{-\partial_a {\cal F}_{\rm inst} } \nn\\
 \gamma_{\rm tot} &=& \gamma \, e^{-2\pi {\rm i} a} \,\frac{\Gamma ({-}2 a)^2 \Gamma \left({-}\frac{3}{2}{+}a{-}\frac{x}{2}\right) \Gamma
		\left(\frac{1}{2}{+}a{-}\frac{x}{2}\right) \Gamma
		\left(\frac{5}{2}{+}a{-}\frac{x}{2}\right)}{\Gamma (2 a)^2 \Gamma
		\left(-\frac{3}{2}{-}a{-}\frac{x}{2}\right) \Gamma \left(\frac{1}{2}{-}a{-}\frac{x}{2}\right)
		\Gamma \left(\frac{5}{2}{-}a{-}\frac{x}{2}\right)}
\eeaq
    The coefficients $c_{ij}, \widehat{c}_{ij}$ in the polynomials
  \be
P_{0 }( y)  = 1+\sum_{i=1}^k   \sum_{j=0}^{i-1}   c_{ij} \, x^i y^{j-i}  \quad , \quad  \widehat{P}_{0 }( y)   = \sum_{i=1}^k   \sum_{j=0}^{i-1}   \widehat{c}_{ij} \, x^i y^{j-i} \\  \label{pp0bis2}
\ee
 are determined by the recursion relations
 \beaq
 c_{ij}&=& \left[  A_1 \,c_{i,j-1}+A_2\, c_{i-1,j}+ A_3\,  c_{i-1,j-1}+\widehat{A}_{1} \widehat{c}_{i,j-1}+\widehat{A}_{2}
 \widehat{c}_{i-1,j} +\widehat{A}_{3} \widehat{c}_{i-1,j-1} \right. \nn\\
 && \left.    +  \sum _{s=1}^{k}
 c_{s,s-1} (C_1  \,  c_{i-s,j-s} + \widehat{C}_1\,  \widehat{c}_{i-s,j-s}) \right]  \left[ 8 \Delta_{ij} (\Delta_{ij}^2 -4 a^2)  \right]^{-1}  \nn\\
 \widehat{c}_{ij}&=& \left[  B_1 \,c_{i,j-1}+B_2\, c_{i-1,j}+ B_3\,  c_{i-1,j-1}+\widehat{B}_{1} \widehat{c}_{i,j-1}+\widehat{B}_{2}
 \widehat{c}_{i-1,j} +\widehat{B}_{3} \widehat{c}_{i-1,j-1} \right. \nn\\
 && \left.    +  \sum _{s=1}^{k}
 c_{s,s-1} (C_2  \,  c_{i-s,j-s} + \widehat{C}_2\,  \widehat{c}_{i-s,j-s}) \right]  \left[ 8 \Delta_{ij} (\Delta_{ij}^2 -4 a^2)  \right]^{-1} \nn\\
 u_i  &=& c_{i,i-1}   \label{csol}
 \eeaq
starting from  $c_{0i}=\delta_{i0}$,  $\widehat{c}_{0i}=0$, with
\beaq
A_1 &=&  -8 \left(\Delta _{ij}+1\right) \left(\Delta _{ij}-x-5\right) ~, ~ \widehat{A}_{1}= 4 \left(4 a^2-(x+5)^2\right) \Delta
_{i,j} \nn\\
A_3 &=&  8 \Delta _{ij}
\left(\Delta _{ij}-x-5\right)~,~ \widehat{A}_{3}= 4 \left(4 a^2-(x+5)^2\right) \left(-\Delta
_{i,j}+x-2\right)\nn\\
A_2&=&  2\left(3 x^2{+}6 x{-}12 a^2{-}13\right) \Delta _{ij}{+}8 \Delta _{ij}^3{-}8 (x{-}1)
\Delta _{ij}^2{+}8 a^2 (x{-}7){-}2( x^3{+}3 x^2{-}13 x{-}15), \nn\\
\widehat{A}_{2}&=&  \left(4 a^2-(x+5)^2\right) \left(-2 (x-1) \Delta _{ij}+4 a^2+x^2-2
x-3\right) \nn\\
B_1&=&  -16 \left(\Delta _{ij}+1\right)~,~ \widehat{B}_{1}= 8 \Delta _{ij}
\left(\Delta _{ij}+x+5\right) ~,~ \widehat{B}_{3}= 8 \left(-\Delta
_{i,j}^2-7 \Delta _{ij}+x^2+3 x-10\right) \nn\\
B_2&=&  4 \left(-2 (x-1) \Delta
_{i,j}+4 a^2+x^2-2 x-3\right) ~,~ B_3= 16 \Delta _{ij} \nn\\
\widehat{B}_{2} &=&  2 \left(-\Delta _{ij}+x+1\right) \left(-4
\Delta _{ij}^2-8 \Delta _{ij}+12 a^2+x^2+2 x-15\right)  \nn\\
C_1 &=&  8 \left(\Delta _{ij}-x-5\right) ~,~ \widehat{C}_1= 16 a^2-4(x+5)^2~,~C_2=  16 ~,~ \widehat{C}_2=8 \left(\Delta
_{ij}+x+5\right)
\eeaq
and $\Delta_{ij}=i-j$.

 Finally one has to compute $a(\ell)$. Given $u_i(a)$ by (\ref{csol}), one can invert the series \eqref{defsigma} to compute $a(u)$ and then set (from \eqref{rdict})
 \be
 u=(\ell+\ft12)^2+\ft{5x}{2} +\ft{x^2}{4}
 \ee
 Inverting the series at high PM order can be time consuming for a Mathematica code. There is an alternative, more direct way to find the function $a(u)$ based on the connection between this quantity and the quantum SW period of a SU(2) theory with three fundamental flavours.
  The quantum SW period $a$ is known to satisfy the continuous fraction equation \cite{Poghosyan:2020zzg}
 \beaq
 \label{fractionequality}
 \frac{ x M(a+1)}{P(a+1) -\frac{ x M(a+2)}{P(a+2)-\ldots
 }}
 +\frac{ x M(a)}{P(a-1)-\frac{ x M(a-1)}{P(a-2)-\ldots
 }}-P(a)=0
 \eeaq
 with
 \beaq
 P(a)&=&a^2+a x+2x -\frac{3 x^2}{4}x-(\ell+\ft12)^2  , \nn\\
   M(a)&=& ( a-\ft{x}{2}-\ft12) ( a-\ft{x}{2}+\ft32) ( a+\ft{x}{2}+\ft32)
 \eeaq
 Equation (\ref{fractionequality}) can be solved for $a(\ell)$ order by order in $x$. For example for $\ell=2,3$ one finds
 \beaq\label{acycleexp}
 a(2) &=&\ft{5}{2}+\ft{107 x^2}{840}-\ft{1695233 x^4}{148176000}+\ft{76720109901233 x^6}{30764716012800000}-\ft{71638806585865707261481
 	x^8}{99644321084605000704000000}+\ldots \nn\\
 a(3) &=&\ft{7}{2}+\ft{13 x^2}{168}-\ft{10921 x^4}{4346496}+\ft{95353832269 x^6}{493385539046400}-\ft{23627105510827613
 	x^8}{1148838359958761472000}+\ldots
 \eeaq

\subsection{The luminosity}

 In this subsection, we collect the formulae needed to evaluate the luminosity at infinity for the case of a particle moving along a circular orbit in a Schwarzschild geometry. \\
   After expanding in harmonics, the corresponding waveform measured by a far away observer sitting at a point with coordinates   $(T,R,\Theta,\Phi)$ can be written as
 \be
 h=  h_{+} -{\rm i}  h_{\times}  \underset{R\to \infty}{\approx} - {4 G\over R} \int  {d\omega\over 2 \pi} \, \sum_{\ell,m} e^{-{\rm i}\omega (T-R_*)}       \, {Z_{\ell m }(\omega) \over 2({\rm i} \omega)^2}    Y_{-2}^{\ell m} (\Theta,\Phi)\;,
 \ee
with $R_*$ the Tortoise coordinate.
 The harmonic coefficients $Z_{\ell m }$ can be written in terms of derivatives of the incoming solution
  \beaq \label{zlmexpr}
Z^{\rm}_{\ell m} &=&  \mathfrak{R}_{\rm in}(y)  \left(\frac{E x^2 b_{0\ell m}}{2 M^2 y (x{-}y)}{+}\frac{ {\rm i} J x^3 \left(y^2{+}4 y{-}4 x\right) b_{1 \ell m}}{8 M^3 y^3 (x{-}y)}{+}\frac{J^2  x^4 \left( 2x{-}4 y{-} y^2\right) b_{2 \ell m}}{64 E M^4 y^3 (x{-}y)}\right)\\
&&+y \partial_y \mathfrak{R}_{\rm in}(y) \left(\frac{ {\rm i} J x^3 b_{1 \ell m}}{4 M^3 y^3}+\frac{J^2 x^4 \left(2x-2 y-2
	y^2\right) b_{2 \ell m}}{16 E M^4 y^5}\right)  {-}y^2 \partial_y^2\mathfrak{R}_{\rm in}(y) \frac{J^2x^4   (x{-}y)  b_{2\ell m}}{16 E M^4 y^5} \nn \eeaq
with
\be
b^0_{\ell m}  ={  \pi   \over 2   }   \sqrt{(\ell^2{-}1)\ell (\ell{+}2)} \,   Y_0^{\ell m}(\ft{\pi}{2},0) ,~~
b^1_{\ell m}  =     \pi     \sqrt{(\ell{-}1)(\ell{+}2) }   Y_{-1}^{\ell m} (\ft{\pi}{2},0),~~
b^2_{\ell m}  =   \pi   \, Y_{-2}^{\ell m} (\ft{\pi}{2},0) \label{bbb} ,
\ee
and
\be
{}_s Y_{\ell m}(\theta,\phi)=e^{i m \phi } \sin ^{2 \ell}\left(\frac{\theta }{2}\right)  \sqrt{\frac{(2 \ell{+}1) (\ell{-}m)! (\ell{+}m)!}{4\pi (\ell{-}s)! (\ell{+}s)!}}
\sum_{r=0}^{\ell{-}s}  ({-})^{\ell{+}m{-}r{-}s} (^{\ell{-}s}_s) (^{\ell+s}_{r{+}s{-}m}) \cot (\ft{\theta}{2})^{2r{+}s{-}m} \label{ylm}
\ee
the spin-weighted spherical harmonics and
\be
   E= {\mu(1-2 v^2)\over \sqrt{1-3 v^2} } \qquad, \qquad  \omega J={\mu  m v^2  \over    \sqrt{1-3 v^2} }
 \ee
  the energy and angular momentum of the light particle.
 Finally, the luminosity,  i.e. the amount of energy per unit time emitted by the system towards infinity,   is computed by the formula (see \cite{Mino:1997bx} and references therein)
\be \label{energy flux}
 {d{\cal E}  \over dt} =\mu^2 \sum_{\ell=2}^\infty \sum_{m=1}^\ell  { |Z^{\rm}_{\ell m}|^2\over 2\pi \omega^2}=\left({d{\cal E}  \over dt}\right)_N  \sum_{\ell=2}^\infty \sum_{m=1}^\ell  \eta_{\ell m}(v)
\ee
where $\mu$ is the mass of the orbiting body.
   Rather than the luminosity, we will display its ratio against the luminosity computed in the Newton approximation
    \be \label{energy newton}
  \left({d{\cal E}\over dt}\right)_N = \frac{32 \mu^2 v^{10}}{5 M^2}
 \ee

\subsection{Numerical tests}

\begin{figure}
    \centering
    \includegraphics[width=0.49\textwidth]{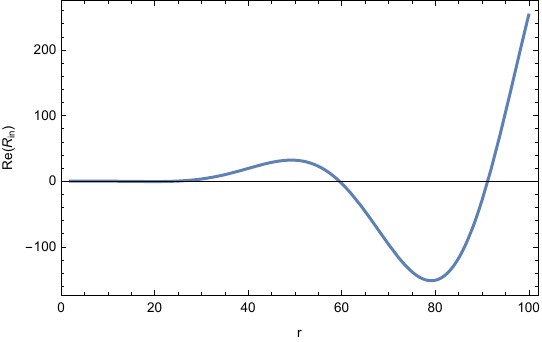}
    \includegraphics[width=0.49\textwidth]{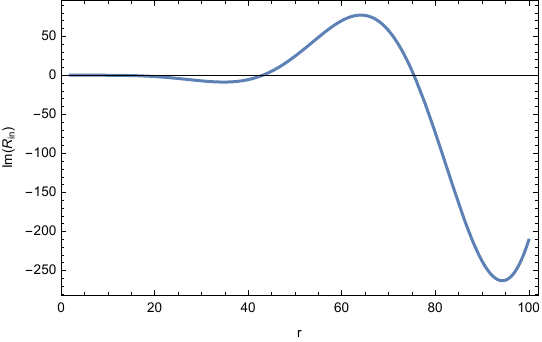}\\
      \includegraphics[width=0.49\textwidth]{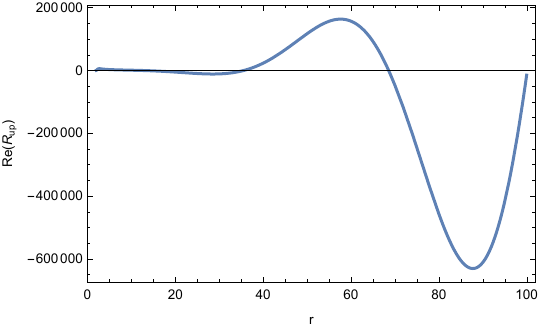}
    \includegraphics[width=0.49\textwidth]{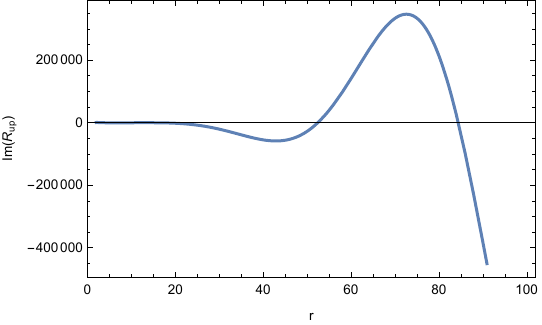}
    \label{fig:immagine1}
    \caption{Plots of incoming and outgoing solutions for $\omega = 0.1, \ell = 2$. 1a: Re$(\mathfrak{R}_{\text{in}})$,  1b: Im$(\mathfrak{R}_{\text{in}})$,
    2a: Re$(\mathfrak{R}_{\text{up}})$,  2b: Im$(\mathfrak{R}_{\text{up}})$.}
    \label{plotL}
\end{figure}

In this subsection we use our formulae computed at order 30PM and set $M=1$. In Figure \ref{plotL}, we plot the real and imaginary parts of the incoming and outgoing solutions obtained with this method for $\ell=2$, $\omega=0.1$.
%
%

The results are compared in Table \ref{trinrup}  against those obtained by the MST method and numerical integration using the BHTK \cite{BHPToolkit}. In all the subsequent tables, the displayed digits are those ones that are insensible to the contribution of terms of order higher than $v^{60}$ in $P_0(y)$ and $\widehat{P}_0(y)$. The agreement is up to 10 and 14 digits even for small values of $r$,  where the PM approximation is less reliable.

\begin{table*}
\centering			
\begin{tabular}{ccccc}
\hline
 $ \mathfrak{R}_{\text{in}}$   & {\small $r=5$ } &  {\small  $r=100$ } \\ \hline
{\small HYP}& {\small -0.004086908784 -{\rm i} 0.000826769654} & {\small 252.949132499007 -{\rm i} 211.392731187846}  \\
{\small BHTK}& {\small -0.004086908784 -{\rm i} 0.000826769654} & {\small 252.949132499009 -{\rm i} 211.392731187848} \\ \hline
$ \mathfrak{R}_{\text{up}}$  &  {\small $r=5$} &  {\small  $r=100$ } \\ \hline
{\small HYP}& {\small 2632.048432 -{\rm i} 840.387207} & {\small  -15205.329855167 -{\rm i} 989985.766975648 }\\
{\small BHTK}& {\small  2632.048432 -{\rm i} 840.387207} &  {\small -15205.329855168 -{\rm i} 989985.766975651 }\\ \hline
\end{tabular}
\caption{Comparison between the 30PM representation of confluent Heun solutions and  MST/Numerical results for $\omega=0.1, \ell=2$. }\label{trinrup}
\end{table*}

\begin{figure}[t]
    \centering
    \includegraphics[width=0.3\textwidth]{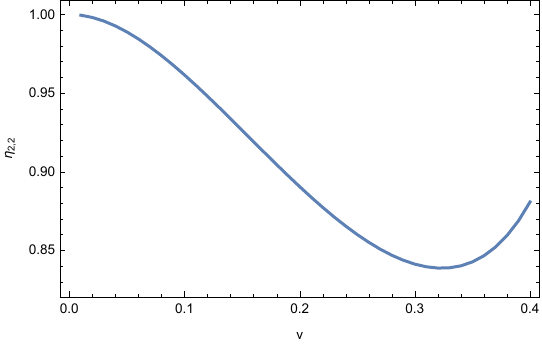}
    \includegraphics[width=0.3\textwidth]{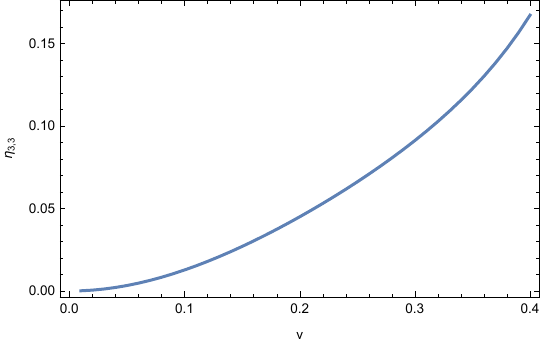}
    \includegraphics[width=0.3\textwidth]{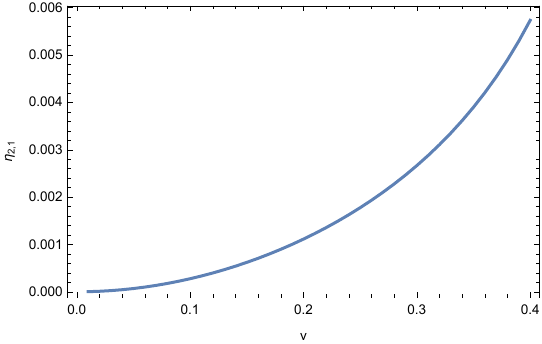}
    \caption{Luminosity vs velocity. We display the $v$-dependence of the ratio $\eta_{\ell m}/\eta^0_{22}$ for the three dominating modes. a: (2,2), b: (3,3), c: (2,1).}
    \label{plotlum}
\end{figure}

In Figure \ref{plotlum} we display the $v$-dependence of the luminosity for the  $(\ell,m)=(2,2)$, $(3,3)$, $(2,1)$ harmonic modes computed from (\ref{rgg}) at order $k=30$  in the PN approximation.

The convergence speed of the results  for some  $(\ell,m)$ modes is showed in Tables \ref{Tconvergence} and \ref{Tconvergence2}.  The digits in bold style underline the  agreement between computations based on the PM hypergeometric representations and the MST/numerical results. As expected, the speed of convergence varies with $v$: in the case of the $(2,2)$ mode, for $v=0.1$ an agreement at 12 digits is already reached at 10PM order, while for $v=0.4$, the contributions at order 30PM  are needed to reach a precision at all the considered digits. A similar behaviour is also found for higher $(\ell,m)$ modes. The last line of each sub-table shows the results coming from the  22PN computation in \cite{Fujita:2012cm}.  We observe that, opposite to what happens if one uses the 22PN expansion, our results agree against those obtained by  the MST/Numerical methods for the first 10-12 digits independently of the $\ell$ mode. The difference is evident in the case of large $\ell$, see for example the results for the $(10,1)$ mode in Table \ref{Tconvergence2}.

\begin{table*}
\centering
\begin{tabular}{|cccc|}
\hline
 $k$ & $\eta_{2,2}(0.1) $ &   $\eta_{2,2}(0.3)$ & $\eta_{2,2}(0.4)$ \\ \hline
5 & \textbf{0.9614837055}13    & \textbf{0.8411}81503167   & \textbf{0.8807}96957664 \\
 10 & \textbf{0.961483705502}  & \textbf{0.84117617}8745 & \textbf{0.8807}13162076 \\
 15 & \textbf{0.961483705502}    & \textbf{0.84117617132}8 & \textbf{0.8807054}81610 \\
 20 & \textbf{0.961483705502}   & \textbf{0.841176171325}& \textbf{0.88070541}2127 \\
 25 & \textbf{0.961483705502}    & \textbf{0.841176171325}  & \textbf{0.8807054116}30 \\
 30 & \textbf{0.961483705502}    & \textbf{0.841176171325 } & \textbf{0.880705411627} \\ \hline
 BHTK& \textbf{0.961483705502}   & \textbf{0.841176171325} & \textbf{0.880705411627} \\  \hline
 22PN& \textbf{0.961483705502}   & \textbf{0.84117617132}6  & \textbf{0.880705}608981  \\ \hline \hline
  $k$ & $  \eta_{2,1}(0.1) $ &   $  \eta_{2,1}(0.3)$ & $  \eta_{2,1}(0.4)$ \\ \hline
  5 & \textbf{0.0002761309188}56    & \textbf{0.0026641}1244976   & \textbf{0.0057}2740466492 \\
 10 & \textbf{0.000276130918862}  & \textbf{0.002664146}79182 & \textbf{0.005730}06836479 \\
 15 & \textbf{0.000276130918862}    & \textbf{0.002664146863}48 & \textbf{0.00573016}433854 \\
 20 & \textbf{0.000276130918862}   & \textbf{0.00266414686353} & \textbf{0.0057301655}1089 \\
 25 & \textbf{0.000276130918862}    & \textbf{0.00266414686353}  & \textbf{0.005730165520}70 \\
 30 & \textbf{0.000276130918862}    & \textbf{0.00266414686353} & \textbf{0.00573016552076} \\  \hline
 BHTK& \textbf{0.000276130918862}   & \textbf{0.00266414686353} & \textbf{0.00573016552076} \\ \hline
 22PN& \textbf{0.000276130918862}   & \textbf{0.00266414686353} & \textbf{0.00573016504468} \\ \hline \hline
  $k$ & $  \eta_{3,3}(0.1) $ &   $  \eta_{3,3}(0.3)$ & $  \eta_{3,3}(0.4)$ \\ \hline
  5 & \textbf{0.0127391414224}    & \textbf{0.091120}2006109   & \textbf{0.16}6992708659 \\
 10 & \textbf{0.0127391414224}  & \textbf{0.0911203195}830 & \textbf{0.167029}390871\\
 15 & \textbf{0.0127391414224}    & \textbf{0.091120319524}5 & \textbf{0.1670292}60059 \\
 20 & \textbf{0.0127391414224}   & \textbf{0.0911203195244} & \textbf{0.1670292582}39 \\
 25 & \textbf{0.0127391414224}    & \textbf{0.0911203195244}  & \textbf{0.167029258222} \\
 30 & \textbf{0.0127391414224}    & \textbf{0.0911203195244} & \textbf{0.167029258222} \\  \hline
 BHTK&\textbf{0.0127391414224}    & \textbf{0.0911203195244}  & \textbf{0.167029258222} \\ \hline
 22PN&\textbf{0.0127391414224}    & \textbf{0.09112031952}62  & \textbf{0.167029}722596  \\ \hline \hline
 $k$ & $ 10^6\times \eta_{3,2}(0.1) $ &   $ 10^4\times \eta_{3,2}(0.3)$ & $ 10^3\times  \eta_{3,2}(0.4)$ \\ \hline
  5 & \textbf{7.695632933} & \textbf{5.99470}001058   & \textbf{2.2415}1427554 \\
 10 & \textbf{7.695632933} & \textbf{5.9947011}1376 & \textbf{2.24157}466375\\
 15 & \textbf{7.695632933}   & \textbf{5.99470112089} & \textbf{2.2415769}1351 \\
 20 & \textbf{7.695632933}   & \textbf{5.99470112089} & \textbf{2.24157695}298\\
 25 & \textbf{7.695632933}    & \textbf{5.99470112089}  & \textbf{2.24157695337} \\
 30 & \textbf{7.695632933}    & \textbf{5.99470112089} & \textbf{2.24157695337} \\ \hline
 BHTK& \textbf{7.695632933}   & \textbf{5.99470112089} & \textbf{2.24157695337} \\ \hline
 22PN& \textbf{7.695632933}    & \textbf{5.9947011208}8  & \textbf{2.241576}33248\ \\ \hline \hline
 $k$ & $10^6 \times \eta_{3,1}(0.1) $ &   $10^6 \times \eta_{3,1}(0.3)$ & $10^5 \times \eta_{3,1}(0.4)$ \\ \hline
5 & \textbf{1.1829878889} & \textbf{8.212183}6172   & \textbf{1.328}87186107 \\
 10 & \textbf{1.1829878889} & \textbf{8.21218325}69 & \textbf{1.32886}658816\\
 15 & \textbf{1.1829878889}   & \textbf{8.2121832518} & \textbf{1.3288658}4761 \\
 20 & \textbf{1.1829878889}   & \textbf{8.2121832518}& \textbf{1.328865835}99\\
 25 & \textbf{1.1829878889}    & \textbf{8.2121832518}  & \textbf{1.32886583589} \\
 30 & \textbf{1.1829878889}    & \textbf{8.2121832518} & \textbf{1.32886583589} \\ \hline
 BHTK& \textbf{1.1829878889}   & \textbf{8.2121832518} & \textbf{1.32886583589} \\ \hline
 22PN& \textbf{1.1829878889}   & \textbf{8.212183251}7  & \textbf{1.328865}77283 \\ \hline \hline
\end{tabular}
\caption{Convergence of the PM$_k$ results for $\eta_{\ell,m}$, with $\ell=2,3, 1 \leq m \leq \ell$, normalized with the Newtonian luminosity.  The results obtained using the MST/ NUM methods and the 22PN computation in \cite{Fujita:2012cm} are displayed in the last two lines of each sub-table.}
\label{Tconvergence}
\end{table*}

 \begin{table*}
\centering
\begin{tabular}{|cccc|}
\hline
 $k$ & $10^{39} \times \eta_{10,1}(0.1) $ &   $10^{31} \times \eta_{10,1}(0.3)$ & $10^{29} \times \eta_{10,1}(0.4)$ \\ \hline
5 & \textbf{9.351820404}3 & \textbf{8.6461}6460005   & \textbf{4.13}218882098\\
 10 & \textbf{9.3518204044} & \textbf{8.64619838996} & \textbf{4.13187842}500\\
 15 & \textbf{9.3518204044}   & \textbf{8.64619838996} & \textbf{4.13187842498}\\
 20 & \textbf{9.3518204044}   & \textbf{8.64619838996}& \textbf{4.13187842498}\\
 25 & \textbf{9.3518204044}    & \textbf{8.64619838996}  & \textbf{4.13187842498} \\
 30 & \textbf{9.3518204044}   & \textbf{8.64619838996} & \textbf{4.13187842498} \\ \hline
 BHTK& \textbf{9.3518204044}   & \textbf{8.64619838996} & \textbf{4.13187842498} \\ \hline
 22PN  & 9.35182040436    & 8.64619786463  & 4.12989060363 \\ \hline
\end{tabular}
\caption{Convergence of the PM$_k$ results for $\eta_{10,1}$ normalized by the Newtonian luminosity.  The results obtained using the MST/ NUM methods and the 22PN computation in \cite{Fujita:2012cm} are displayed in the last two lines.}
\label{Tconvergence2}
\end{table*}

Finally Table \ref{comparison dE/dt} we compare our 30PN results for the energy flux against those obtained by MST/Numerical methods and  by the  22PN computation in \cite{Fujita:2012cm}.

\begin{table*}
\centering
\begin{tabular}{|cccc|}
\hline
 & $v=0.1$   & $v=0.3$ & $v=0.4$ \\ \hline
HYP & 0.974719282291879  & 0.950003411484245 & 1.10991197121 \\ \hline
BHTK & 0.974719282291999   & 0.950003411484254 & 1.10991197120 \\ \hline
22PN  & 0.974719282291877   & 0.950003411477562 & 1.10990619336 \\ \hline
\end{tabular}
\caption{Total luminosity  $d{\cal E}/dt$.  }\label{comparison dE/dt}
\end{table*}

We notice that in the whole range of $v$ the results obtained with the 30PN hypergeometric representations for the total luminosity $d{\cal E}/dt $ agree with those obtained by the numerical integration with the MST method up to 11-13 digits.

\subsection*{Mathematica codes}
Two Mathematica notebooks are given as ancillary files:
\begin{itemize}
\item  \textbf{RinRupdRdt.nb} With this file it is possible to compute the $\eta_{\ell m}$ modes importing the data  files  P30.m, Pt30.m, gg30.m, gk30.m and a30.m, containing the polynomials $P_0(y)$, $\widehat{P}_0(y)$, the quantities $\gamma$, $\gamma_\kappa$ and the $a$ cycle. The results thus obtained will show a slower convergence with respect to the data of Tables \ref{trinrup}, \ref{Tconvergence}-\ref{Tconvergence2} and \ref{comparison dE/dt} which were obtained keeping higher orders than 30PN for
    $c_{ij}, \widehat{c}_{ij}$.  The files with these data are quite large and can be obtained by writing to any of the authors.

\item   \textbf{PNexpansions.nb}  computes the PN expansions of $R_+(y)$, $\gamma$, $\gamma_\kappa$ at any PM order lower than 10. This code imports the
 data  files  PNRpp10.m,  PNgg10.m, PNgk10.m and PNaa10.m, containing the expansions of the quantities  $R_+(y)$, $\gamma$, $\gamma_\kappa$ and $a$ at order 10PM.
 If these files are used to reproduce our results  above the convergence will be  a little slower

 \end{itemize}
\section{Summary and conclusions}\label{sec6}

  In this paper, using black hole perturbation theory, we derive exact formulae for the PM$_k$ and PN$_k$ expansions (at order $k$ in the soft-limit $\omega M \ll 1$) of the gravitational wave form
  emitted by a particle moving on an arbitrary trajectory in Schwarzschild geometry at leading order in the probe limit. The results rely on a novel
    hypergeometric representation of the solutions of the Heun (or confluent Heun) equation (see Appendix \ref{sec5} for the discussion of the non-confluence case).  We build four equivalent basis of solutions, $\widetilde{G}^1_\alpha$,  $G^1_\alpha$, $G^0_\alpha$ and $\widetilde{G}^0_\alpha$ that describe with good accuracy different regions of the space time, and are connected to each other by Heun connection matrices.
   The  final picture is summarized in the following scheme
   \beaq
\begin{array}{lll}
  {\rm Near~horizon} \qquad  &    {\rm Near~zone}  \qquad\qquad  &   {\rm Wave~ zone}   \\
 z\approx 1 &  x\ll z \ll 1  & z \ll x \ll 1   \\
 F_{\alpha \alpha'} \widetilde{G}^1_{\alpha'} & G^1_\alpha =g_\alpha   G^0_\alpha  &   g_\alpha B_{\alpha \alpha'} \widetilde{G}^0_{\alpha'}
\end{array}
\eeaq
 where $B_{\alpha \alpha'}$ and $F_{\alpha \alpha'}$ are the standard hypergeometric connection matrices relating $z$ to $1/z$ or $1-z$ respectively. The whole non-triviality of the Heun connection formula is codified  in the dependence of these matrices on  the characteristic function $a(\ell)$. This simple result is a direct consequence of our ansatz for the solution of the differential equation in terms  of a single hypergeometric function and its derivative.

   Specifying to circular orbits, we derive the probe limit of the wave form and energy flux at order 30PN.  In contrast with the PN expansions that one can find in literature \cite{Blanchet:2013haa,Fujita:2012cm}, these formulae  resum the infinite tower of logarithmically divergent terms (tail and tail of tails), responsible for the slow convergence of the PN series, into exponentials. Given that the results are too large to appear in print we collect them in ancillary mathematica files. The results for the waveform have been compared against those obtained using the numerical package BHToolkit, finding an agreement from 9 to 12 digits in the whole range of parameters allowed for bounded orbits.

   We remark that our results provide a hypergeometric representations of solutions of Heun equations of general type, so they apply to any
    physical problem governed by equations of Heun type. In particular, they can be easily adapted to the study of collisions of spinning BHs (Kerr geometry) or orbital motions in the backgrounds of topological stars, D-branes and cosmological perturbations \cite{Bianchi:2021xpr,Bianchi:2021mft,DiRusso:2024hmd,Bianchi:2024vmi,Bianchi:2024mlq,Bianchi:2024rod}. It would also be very interesting to apply our results to the generation of templates of waveforms for GWs data analysis\footnote{We thank A.Nagar and M.Panzeri for useful discussions on this issue.}. We hope to also come back to this issue in a future publication.

\section*{Acknowledgements}
F.~F. and J.~F.~M.  were partially supported by the MIUR PRIN contract 2020KR4KN2 ``String Theory as a bridge between Gauge Theories and Quantum Gravity'' and the INFN project ST\&FI ``String Theory and Fundamental Interactions''. Moreover they thank M. Bianchi, M. Billò, D. Bini, G. Bonelli, M.L. Frau, A.Lerda, A.Nagar, M.Panzeri, A. Tanzini for fruitful scientific exchanges. A.~C. would like to thank the Department of Theoretical Physics at CERN for kind hospitality during the final stages of the present work. The research of R.~P. and
supported by the Armenian SCS  grants 21AG-1C060 and 24WS-1C031.
H.~P. received support from the Armenian SCS  grants  21AG-1C062 and 24WS-1C031.

 \begin{appendix}

\section{The Heun equation}\label{sec5}

 The generalization of the previous results to the case of the non-confluent Heun equation is straightforward.
The Heun differential equation is defined as
\begin{equation}
	\label{Heun}
	G_1''(z) + \left( {\beta_0\over z} +{\beta_1\over z-1}+{\beta_2\over z-\lambda} \right) G_1'(z) +{ z \alpha_0 \alpha_1 - \sigma \over z (z-1)(z-\lambda)}  G_1(z)=0
\end{equation}
with  $\beta_2=1+\alpha_0+\alpha_1-\beta_0-\beta_1$.  The solutions of the Heun equation can be written as linear combinations of
\beaq
{\rm Heun}_-(z) &=& {\rm Heun}\left[\lambda,\sigma ,\alpha_0 ,\alpha_1 ,\beta _0,\beta _1,z\right]\\
{\rm Heun}_+(z) &=&  z^{1{-}\beta _0} {\rm Heun}\left[\lambda,\sigma {+}\left(1{-}\beta_0 \right) \left(  \beta_2{+}\lambda \beta_1  \right),1{+}\alpha_0 {-}\beta _0,1{+}\alpha_1 {-}\beta _0,2{-}\beta_0,\beta _1,z\right]\nn
\eeaq
where ${\rm Heun}\left[\lambda,\sigma ,\alpha_0 ,\alpha_1 ,\beta _0,\beta _1,z\right]$ is the Heun function which is normalized to one at $z=0$, so that
 \be
 {\rm Heun}_-(z)=1+\ldots \,, \, {\rm Heun}_+(z)=z^{1-\beta_0} (1+\ldots )
 \ee
The dots stand for corrections in $z$. The expansion coefficients of ${\rm Heun}_\alpha(z)$ can be obtained recursively, solving the equation  order by order in $z$.
The Heun differential equation can be related to the quantum SW curve of a ${\cal N}=2$ supersymmetric gauge theory with four fundamental hypermultiplets.
To this aim, we write the six Heun parameters in terms of 4 masses $m_i$, a gauge coupling $x$ and a Coulomb branch parameter $u$, defined via the identifications\footnote{
Equation (\ref{Heun}) can be alternatively written in the normal form (\ref{normal}) with
\be
G_1(z) =  (1-z)^{-\frac{\beta _1}{2}} z^{-\frac{\beta _0}{2}}  (z-\lambda )^{-\frac{\beta_2}{2}} \, \Psi (z)
\ee
and
\beaq
Q_{\rm gauge} &=& \frac{\ft{1}{4}-
   \left(\ft{m_3+m_4}{2}\right){}^2}{(z-x)^2}+\frac{\ft{1}{4}-
   \left( \ft{m_3-m_4}{2}\right){}^2}{z^2}+\frac{\ft{1}{4}-
   \left(\ft{m_1+m_2}{2}\right){}^2}{(z-1)^2}+\frac{m_3^2+m_4^2+2 m_1
   m_2-1}{2 (z-1) z} \nn\\
 &&+  \frac{(1-x) \left( u+\ft{1}{4} -\frac{m_3^2}{2}-\frac{m_4^2}{2}\right)
    -\ft{x}{2}\left(1{-}m_1{-}m_2\right)
   \left(1{-}m_3{-}m_4\right)  }{
   (z-1) z (z-x)}
\eeaq
}
\beaq
  \alpha _0&=& 1{-}m_1{-}m_3 \quad ,\quad \alpha _1= 1{-}m_2{-}m_3\quad , \quad    \lambda =x  \\
  \beta_0&=& 1{-}m_3{+}m_4\quad ,\quad \beta _1=1 {-}m_1{-}m_2 \quad , \quad    \beta_2 =1-m_3-m_4  \nn\\
  \sigma &=&u (x-1)+\left(\ft{1}{2}-m_3\right){}^2+x \left(\ft34{-}m_1{-} m_2{-} m_3{+}m_1 m_2{+}m_1 m_3{+}m_2m_3  \right) \nn\eeaq
The mathematical problem we are interested in is looking for a solution in an interval, let us say   $z\in [0,1]$, satisfying some specific boundary conditions.
The general solution
\be
G_1(z)=\sum_\alpha B_\alpha \, {\rm Heun}_\alpha(z)
\ee
behaves, near the boundaries, as
\beaq
G_1(z) &\underset{z\to 1}{\approx}&  D_{-}(1+\ldots)  +D_{+} (1{-}z)^{m_1+m_2} (1+\ldots)
  \nn\\
 G_1(z) &\underset{z\to 0}{\approx}&  B_{-} (1+\ldots) +B_{+} z^{m_3-m_4} (1+\ldots)   \label{gbcgen}
\eeaq
Specifying boundary conditions corresponds to fix two linear combinations made out of the four coefficients $B_{\pm}$,  $D_{\pm}$.

\subsection{The hypergeometric representation}

 As before we start from the PM/PN ans\"atze
\be
   G^0_\alpha(\ft{x}{z} ) g_\alpha(x)
    =   G^1_\alpha(z )+\ldots   \label{gp0}
\ee
with
\beaq
 G^0_\alpha (y)  &=&      P_{0}( y) \, H^0_\alpha (y) {+} \widehat{P}_{0}( y) \,  y H^{0}_\alpha{}'(y)  \nn\\
  G^1_\alpha (z) &=&      P_{1}(z) \, H^1_\alpha (z) {+} \widehat{P}_{1}(z) \,  z  H^{1}_\alpha{}'(z )    \label{g0g1h} \nn\\
  H^0_\alpha(y) &=& y^{\frac{1}{2}- \alpha a  - m_3} \,
   _2F_1\left( \ft12- \alpha a -m_3, \ft12- \alpha a
   -m_4; 1-2  \alpha a;y\right) \nn\\
  H^1_\alpha(z) &=& z^{-\frac{1}{2}+  \alpha a  + m_3} \,
   _2F_1\left( \ft12+ \alpha a -m_1,\ft12+ \alpha a
   -m_2;1+2   \alpha a;z\right) \label{ghnonc}
\eeaq
and
\beaq
P_{0 }(y) & = &1+\sum_{i=1}^k   \sum_{j=0}^{i-1}   c_{ij} \, x^i y^{j-i}  \quad ,\quad \widehat{P}_{0 }(y)   = \sum_{i=1}^k   \sum_{j=0}^{i}   \widehat{c}_{ij} \, x^i y^{j-i} \nn\\ \\
P_{1 }(z)  &   = & 1+  \sum_{j=1}^k \sum_{i=0}^{j-1}    d_{ij}   \, z^{i-j} x^j  \quad ,\quad \widehat{P}_{1 }(z) = \sum_{j=1}^k \sum_{i=0}^{j}    \widehat{d}_{ij}   \, z^{i-j} x^j  \nn \label{pp2}
\eeaq
   We notice that now only the non-confluent hypergeometric functions are involved. As before, plugging the ans\"atze (\ref{gp0})  into (\ref{Heun}) and setting to zero the coefficients of $H^p_\alpha$ ($p=0,1$) and its derivatives, one finds a linear system of algebraic equations for the coefficients $c_{ij}$, $\widehat{c}_{ij}$, $d_{ij}$, $\widehat{d}_{ij}$.
  For example, for $k=1$, writing $u=a^2+u_1 x+\ldots$, one finds
\beaq
 c_{10} &=&
\frac{1}{2}(1-m_1-m_2-m_3) -\frac{2 m_1 m_2 m_3}{4 a^2-1}   \quad, \quad   \widehat{c}_{11} =  -\widehat{c}_{10}=\frac{1}{2}+ \frac{2 m_1 m_2}{4a^2-1}  \nn\\
 d_{01} &= &  \frac{2 \left(m_3-1\right) m_3 m_4}{4 a^2-1}-\frac{m_4}{2} \quad ,  \quad
 \widehat{d}_{11} = - \widehat{d}_{01}= \frac{2 m_3 m_4}{4 a^2-1}+\frac{1}{2} \nn\\
u_1 &=&  \frac{2 m_1 m_2 m_3 m_4}{4 a^2-1}+\frac{1}{8} \left(4
  a^2-1\right)+\frac{1}{2} \left( 1 - \sum_{i=1}^4 m_i+ \sum_{i<j}^4 m_i m_j   \right) \label{pp01s}
\eeaq
   As in the confluent case, we denote by
   \be
   g_\alpha(x) ={G^1_\alpha(z)\over G^0_\alpha (\ft{x}{z})}
   \ee
    the ratio of the two solutions.  Explicit computations show that, as in the confluent case, the ratio of the two components
   can be written as
  \be
\gamma(x) ={g_+(x)\over g_-(x) }=   x^{2a} e^{-\partial_a {\cal F}_{\rm inst}(a,x) } =x^{2 a} \left(1+ x \left(a -\frac{16 a m_1 m_2 m_3 m_4 }{\left(1-4 a^2 \right)^2} \right) + \ldots \right)  \label{gammaHeun}
\ee
with  ${\cal F}_{\rm inst} $ still is given by (\ref{finst}).
  The function  ${\cal F}_{\rm inst} $ corresponds now to the prepotential of a ${\cal N}=2$ supersymmetric $SU(2)$ gauge theory with four fundamental flavours.

\subsection{Connection formulae}

 As in the confluent case, connection formulae for the Heun function can be easily derived from the standard hypergeometric transformation rules of $H^p_\alpha$.
 We introduce the two extra sets of hypergeometric functions
{
	\beaq
	\widetilde{H}^{0}_{\alpha}(y) & = &  ( e^{ {-} i \pi }
   y)^{{-}\frac{1{+}\alpha}{2} m_{34} } \,
   _2F_1\left( \ft12{-}a{-}m_3{+}\ft{1{+}\alpha}{2}
  m_{34}, \ft12{+}a{-}m_3{+}\ft{1{+}\alpha}{2} m_{34}; 1{+}\alpha
  m_{34};\ft{1}{y}\right)   \label{hyp2}  \\
	\widetilde{H}^{1}_{\alpha}(z) & = & z^{a{+}m_3{-}\frac{1}{2}} (1{-}z)^{\frac{1+\alpha}{2}
   \left(m_1{+}m_2\right)} \, _2F_1\left( \ft12{+}a{+}\alpha
   m_1,\ft12{+}a{+}\alpha  m_2;1{+}\alpha
   \left(m_1+m_2\right);1{-}z\right)  \nn
	\eeaq
}
with $m_{34}=m_3-m_4$.
We notice that both $\widetilde{H}^0_{\alpha}(y)$ and $\widetilde{H}^1_{\alpha}(z)$ are invariant under the sign flipping $a \to -a$, due to hypergeometric identities.
The functions (\ref{hyp2}) are related to those introduced in (\ref{ghnonc}) by the standard hypergeometric relations
\be
H^0_\alpha (y) = \sum_{\alpha'=\pm}  B_{\alpha \alpha'}  \widetilde{H}^{0}_{\alpha'}(y) \quad , \quad
H^1_\alpha(z) = \sum_{\alpha'=\pm}  F_{\alpha \alpha' }  \widetilde{H}^{1}_{\alpha'}(z)
\label{g0con00}
\ee
with
\beaq
B_{\alpha \alpha'} &=&e^{-i \pi  \left( \alpha a
   +m_3-\frac{1}{2}\right)} \frac{\Gamma (1-2 \alpha a )  \Gamma \left( \alpha'\left(m_4-m_3\right) \right)}{\Gamma \left( \ft12 -\alpha a  - \alpha' m_3 \right)
    \Gamma \left( \ft12 -\alpha a + \alpha' m_4 \right)} \nn\\
F_{\alpha\alpha'}  &=&\frac{\Gamma (1+2 \alpha a ) \Gamma
   \left(-\alpha' \left(m_1+m_2\right) \right)}{\Gamma \left( \ft12 +\alpha a  - \alpha' m_1  \right) \Gamma \left( \ft12 +\alpha a  - \alpha' m_2 \right)}
\eeaq
Plugging  (\ref{g0con00}) into (\ref{g0g1h}) one finds the connection formulae
\beaq
   G^0_\alpha(y) &=&   \sum_{\alpha'=\pm}  B_{\alpha \alpha'}  \widetilde{G}^0_{\alpha'} (y) ,\nn\\
   G^1_\alpha (z) &=& \sum_{\alpha'=\pm}  F_{\alpha \alpha'} \widetilde{G}^1_{\alpha'}(z)    \label{gp21}
\eeaq
with
\beaq
  \widetilde{G}^0_{\alpha}(y) &=&  P_{0}(y)   \widetilde{H}^{0}_{\alpha}(y)   {+} \widehat{P}_{0}(\ft{x}{z})   y \widetilde{H}^{0}_{\alpha}{}'(y)   \nn\\
  \widetilde{G}^1_{\alpha}(z) &=&  P_{1}(z) \widetilde{H}^{ 1}_{\alpha}(z)   {+} \widehat{P}_{1}(z)   z \widetilde{H}^{ 1 }_{\alpha}{}'(z)  \label{gtildes}
\eeaq
describing the solution in the \textit{far region} and \textit{near horizon zone}.
The pairs $ G^0_\alpha(z)$, $ G^1_\alpha(z)$, $\widetilde{G}^0_{\alpha}(z)$, $ \widetilde{G}^0_{\alpha}(z) $ provide four different bases of solutions of the Heun equation.
Each basis covers a different patch of the interval $z\in [0,1]$. They are related to each other by the connection formulae (\ref{gp21}). Remarkably the Heun connection formulae (\ref{gp21}) take the familiar hypergeometric form, with the important difference that the argument of the gamma functions depends on the non-trivial function $a(u)$.
Finally, we notice that the Heun function corresponds to the minus component of the far region basis
\be
   {\rm Heun}_- (z) = \widetilde{G}^0_-(z)  \underset{z\to 0}{\approx} 1+ \ldots \label{heunm}
\ee
Indeed in this limit $P_0\to 1$, $\widetilde{H}^0_- \to 1$ and (\ref{heunm}) follows from (\ref{gtildes}).

 \end{appendix}

  \providecommand{\href}[2]{#2}\begingroup\raggedright\endgroup


\begin{thebibliography}{10}

\bibitem{LISA:2022kgy}
{\scshape LISA} collaboration, \emph{{New horizons for fundamental physics with
  LISA}}, \href{https://doi.org/10.1007/s41114-022-00036-9}{\emph{Living Rev.
  Rel.} {\bfseries 25} (2022) 4}
  [\href{https://arxiv.org/abs/2205.01597}{{\ttfamily 2205.01597}}].

\bibitem{Cardoso:2019rvt}
V.~Cardoso and P.~Pani, \emph{{Testing the nature of dark compact objects: a
  status report}},
  \href{https://doi.org/10.1007/s41114-019-0020-4}{\emph{Living Rev. Rel.}
  {\bfseries 22} (2019) 4} [\href{https://arxiv.org/abs/1904.05363}{{\ttfamily
  1904.05363}}].

\bibitem{Gair:2012nm}
J.R.~Gair, M.~Vallisneri, S.L.~Larson and J.G.~Baker, \emph{{Testing General
  Relativity with Low-Frequency, Space-Based Gravitational-Wave Detectors}},
  \href{https://doi.org/10.12942/lrr-2013-7}{\emph{Living Rev.Rel.} {\bfseries
  16} (2013) 7} [\href{https://arxiv.org/abs/1212.5575}{{\ttfamily
  1212.5575}}].

\bibitem{Yunes:2013dva}
N.~Yunes and X.~Siemens, \emph{{Gravitational-Wave Tests of General Relativity
  with Ground-Based Detectors and Pulsar Timing-Arrays}},
  \href{https://doi.org/10.12942/lrr-2013-9}{\emph{Living Rev.Rel.} {\bfseries
  16} (2013) 9} [\href{https://arxiv.org/abs/1304.3473}{{\ttfamily
  1304.3473}}].

\bibitem{Thorne:1980ru}
K.S.~Thorne, \emph{{Multipole Expansions of Gravitational Radiation}},
  \href{https://doi.org/10.1103/RevModPhys.52.299}{\emph{Rev. Mod. Phys.}
  {\bfseries 52} (1980) 299}.

\bibitem{Damour:1982wm}
T.~Damour, \emph{{GRAVITATIONAL RADIATION AND THE MOTION OF COMPACT BODIES}},
  in \emph{{Les Houches Summer School on Gravitational Radiation}}, 8, 1982.

\bibitem{Blanchet:2013haa}
L.~Blanchet, \emph{{Post-Newtonian Theory for Gravitational Waves}},
  \href{https://doi.org/10.12942/lrr-2014-2}{\emph{Living Rev. Rel.} {\bfseries
  17} (2014) 2} [\href{https://arxiv.org/abs/1310.1528}{{\ttfamily
  1310.1528}}].

\bibitem{Teukolsky:1973ha}
S.A.~Teukolsky, \emph{{Perturbations of a rotating black hole. 1. Fundamental
  equations for gravitational electromagnetic and neutrino field
  perturbations}}, \href{https://doi.org/10.1086/152444}{\emph{Astrophys. J.}
  {\bfseries 185} (1973) 635}.

\bibitem{Poisson:1993vp}
E.~Poisson, \emph{{Gravitational radiation from a particle in circular orbit
  around a black hole. 1: Analytical results for the nonrotating case}},
  \href{https://doi.org/10.1103/PhysRevD.47.1497}{\emph{Phys. Rev. D}
  {\bfseries 47} (1993) 1497}.

\bibitem{Tagoshi:1993dm}
H.~Tagoshi and T.~Nakamura, \emph{{Gravitational waves from a point particle in
  circular orbit around a black hole: Logarithmic terms in the postNewtonian
  expansion}}, \href{https://doi.org/10.1103/PhysRevD.49.4016}{\emph{Phys. Rev.
  D} {\bfseries 49} (1994) 4016}.

\bibitem{Poisson:1994yf}
E.~Poisson and M.~Sasaki, \emph{{Gravitational radiation from a particle in
  circular orbit around a black hole. 5: Black hole absorption and tail
  corrections}}, \href{https://doi.org/10.1103/PhysRevD.51.5753}{\emph{Phys.
  Rev. D} {\bfseries 51} (1995) 5753}
  [\href{https://arxiv.org/abs/gr-qc/9412027}{{\ttfamily gr-qc/9412027}}].

\bibitem{Shibata:1994jx}
M.~Shibata, M.~Sasaki, H.~Tagoshi and T.~Tanaka, \emph{{Gravitational waves
  from a particle orbiting around a rotating black hole: PostNewtonian
  expansion}}, \href{https://doi.org/10.1103/PhysRevD.51.1646}{\emph{Phys. Rev.
  D} {\bfseries 51} (1995) 1646}
  [\href{https://arxiv.org/abs/gr-qc/9409054}{{\ttfamily gr-qc/9409054}}].

\bibitem{Tagoshi:1994sm}
H.~Tagoshi and M.~Sasaki, \emph{{PostNewtonian expansion of gravitational waves
  from a particle in circular orbit around a Schwarzschild black hole}},
  \href{https://doi.org/10.1143/PTP.92.745}{\emph{Prog. Theor. Phys.}
  {\bfseries 92} (1994) 745}
  [\href{https://arxiv.org/abs/gr-qc/9405062}{{\ttfamily gr-qc/9405062}}].

\bibitem{Tanaka:1996lfd}
T.~Tanaka, H.~Tagoshi and M.~Sasaki, \emph{{Gravitational waves by a particle
  in circular orbits around a Schwarzschild black hole: 5.5 postNewtonian
  formula}}, \href{https://doi.org/10.1143/PTP.96.1087}{\emph{Prog. Theor.
  Phys.} {\bfseries 96} (1996) 1087}
  [\href{https://arxiv.org/abs/gr-qc/9701050}{{\ttfamily gr-qc/9701050}}].

\bibitem{Mano:1996vt}
S.~Mano, H.~Suzuki and E.~Takasugi, \emph{{Analytic solutions of the Teukolsky
  equation and their low frequency expansions}},
  \href{https://doi.org/10.1143/PTP.95.1079}{\emph{Prog. Theor. Phys.}
  {\bfseries 95} (1996) 1079}
  [\href{https://arxiv.org/abs/gr-qc/9603020}{{\ttfamily gr-qc/9603020}}].

\bibitem{Mino:1997bx}
Y.~Mino, M.~Sasaki, M.~Shibata, H.~Tagoshi and T.~Tanaka, \emph{{Black hole
  perturbation: Chapter 1}},
  \href{https://doi.org/10.1143/PTPS.128.1}{\emph{Prog. Theor. Phys. Suppl.}
  {\bfseries 128} (1997) 1}
  [\href{https://arxiv.org/abs/gr-qc/9712057}{{\ttfamily gr-qc/9712057}}].

\bibitem{Suzuki:1998vy}
H.~Suzuki, E.~Takasugi and H.~Umetsu, \emph{{Perturbations of Kerr-de Sitter
  black hole and Heun's equations}},
  \href{https://doi.org/10.1143/PTP.100.491}{\emph{Prog. Theor. Phys.}
  {\bfseries 100} (1998) 491}
  [\href{https://arxiv.org/abs/gr-qc/9805064}{{\ttfamily gr-qc/9805064}}].

\bibitem{Fujita:2010xj}
R.~Fujita and B.R.~Iyer, \emph{{Spherical harmonic modes of 5.5 post-Newtonian
  gravitational wave polarisations and associated factorised resummed waveforms
  for a particle in circular orbit around a Schwarzschild black hole}},
  \href{https://doi.org/10.1103/PhysRevD.82.044051}{\emph{Phys. Rev. D}
  {\bfseries 82} (2010) 044051}
  [\href{https://arxiv.org/abs/1005.2266}{{\ttfamily 1005.2266}}].

\bibitem{Fujita:2011zk}
R.~Fujita, \emph{{Gravitational radiation for extreme mass ratio inspirals to
  the 14th post-Newtonian order}},
  \href{https://doi.org/10.1143/PTP.127.583}{\emph{Prog. Theor. Phys.}
  {\bfseries 127} (2012) 583}
  [\href{https://arxiv.org/abs/1104.5615}{{\ttfamily 1104.5615}}].

\bibitem{Fujita:2012cm}
R.~Fujita, \emph{{Gravitational Waves from a Particle in Circular Orbits around
  a Schwarzschild Black Hole to the 22nd Post-Newtonian Order}},
  \href{https://doi.org/10.1143/PTP.128.971}{\emph{Prog. Theor. Phys.}
  {\bfseries 128} (2012) 971}
  [\href{https://arxiv.org/abs/1211.5535}{{\ttfamily 1211.5535}}].

\bibitem{Cutler:1993vq}
C.~Cutler, E.~Poisson, G.J.~Sussman and L.S.~Finn, \emph{{Gravitational
  radiation from a particle in circular orbit around a black hole. 2: Numerical
  results for the nonrotating case}},
  \href{https://doi.org/10.1103/PhysRevD.47.1511}{\emph{Phys. Rev. D}
  {\bfseries 47} (1993) 1511}.

\bibitem{Damour:2008gu}
T.~Damour, B.R.~Iyer and A.~Nagar, \emph{{Improved resummation of
  post-Newtonian multipolar waveforms from circularized compact binaries}},
  \href{https://doi.org/10.1103/PhysRevD.79.064004}{\emph{Phys. Rev. D}
  {\bfseries 79} (2009) 064004}
  [\href{https://arxiv.org/abs/0811.2069}{{\ttfamily 0811.2069}}].

\bibitem{Damour:2009kr}
T.~Damour and A.~Nagar, \emph{{An Improved analytical description of
  inspiralling and coalescing black-hole binaries}},
  \href{https://doi.org/10.1103/PhysRevD.79.081503}{\emph{Phys. Rev. D}
  {\bfseries 79} (2009) 081503}
  [\href{https://arxiv.org/abs/0902.0136}{{\ttfamily 0902.0136}}].

\bibitem{Buonanno:2009zt}
A.~Buonanno, B.~Iyer, E.~Ochsner, Y.~Pan and B.S.~Sathyaprakash,
  \emph{{Comparison of post-Newtonian templates for compact binary inspiral
  signals in gravitational-wave detectors}},
  \href{https://doi.org/10.1103/PhysRevD.80.084043}{\emph{Phys. Rev. D}
  {\bfseries 80} (2009) 084043}
  [\href{https://arxiv.org/abs/0907.0700}{{\ttfamily 0907.0700}}].

\bibitem{Damour:2012ky}
T.~Damour, A.~Nagar and S.~Bernuzzi, \emph{{Improved effective-one-body
  description of coalescing nonspinning black-hole binaries and its
  numerical-relativity completion}},
  \href{https://doi.org/10.1103/PhysRevD.87.084035}{\emph{Phys. Rev. D}
  {\bfseries 87} (2013) 084035}
  [\href{https://arxiv.org/abs/1212.4357}{{\ttfamily 1212.4357}}].

\bibitem{Aminov:2020yma}
G.~Aminov, A.~Grassi and Y.~Hatsuda, \emph{{Black Hole Quasinormal Modes and
  Seiberg-Witten Theory}},  \href{https://arxiv.org/abs/2006.06111}{{\ttfamily
  2006.06111}}.

\bibitem{Bianchi:2021xpr}
M.~Bianchi, D.~Consoli, A.~Grillo and J.F.~Morales, \emph{{QNMs of branes, BHs
  and fuzzballs from Quantum SW geometries}},
  \href{https://arxiv.org/abs/2105.04245}{{\ttfamily 2105.04245}}.

\bibitem{Bianchi:2021mft}
M.~Bianchi, D.~Consoli, A.~Grillo and J.F.~Morales, \emph{{More on the SW-QNM
  correspondence}}, \href{https://doi.org/10.1007/JHEP01(2022)024}{\emph{JHEP}
  {\bfseries 01} (2022) 024}
  [\href{https://arxiv.org/abs/2109.09804}{{\ttfamily 2109.09804}}].

\bibitem{Bonelli:2021uvf}
G.~Bonelli, C.~Iossa, D.P.~Lichtig and A.~Tanzini, \emph{{Exact solution of
  Kerr black hole perturbations via CFT2 and instanton counting: Greybody
  factor, quasinormal modes, and Love numbers}},
  \href{https://doi.org/10.1103/PhysRevD.105.044047}{\emph{Phys. Rev. D}
  {\bfseries 105} (2022) 044047}
  [\href{https://arxiv.org/abs/2105.04483}{{\ttfamily 2105.04483}}].

\bibitem{Bonelli:2022ten}
G.~Bonelli, C.~Iossa, D.P.~Lichtig and A.~Tanzini, \emph{{Irregular Liouville
  correlators and connection formulae for Heun functions}},
  \href{https://arxiv.org/abs/2201.04491}{{\ttfamily 2201.04491}}.

\bibitem{Consoli:2022eey}
D.~Consoli, F.~Fucito, J.F.~Morales and R.~Poghossian, \emph{{CFT description
  of BH\textquoteright{}s and ECO\textquoteright{}s: QNMs, superradiance,
  echoes and tidal responses}},
  \href{https://doi.org/10.1007/JHEP12(2022)115}{\emph{JHEP} {\bfseries 12}
  (2022) 115} [\href{https://arxiv.org/abs/2206.09437}{{\ttfamily
  2206.09437}}].

\bibitem{Bautista:2023sdf}
Y.F.~Bautista, G.~Bonelli, C.~Iossa, A.~Tanzini and Z.~Zhou, \emph{{Black hole
  perturbation theory meets CFT2: Kerr-Compton amplitudes from
  Nekrasov-Shatashvili functions}},
  \href{https://doi.org/10.1103/PhysRevD.109.084071}{\emph{Phys. Rev. D}
  {\bfseries 109} (2024) 084071}
  [\href{https://arxiv.org/abs/2312.05965}{{\ttfamily 2312.05965}}].

\bibitem{Aminov:2023jve}
G.~Aminov, P.~Arnaudo, G.~Bonelli, A.~Grassi and A.~Tanzini, \emph{{Black hole
  perturbation theory and multiple polylogarithms}},
  \href{https://doi.org/10.1007/JHEP11(2023)059}{\emph{JHEP} {\bfseries 11}
  (2023) 059} [\href{https://arxiv.org/abs/2307.10141}{{\ttfamily
  2307.10141}}].

\bibitem{ToVSapQNM}
T.~Ikeda, M.~Bianchi, D.~Consoli, A.~Grillo, J.F.~Morales, P.~Pani et~al.,
  \emph{{Black-hole microstate spectroscopy: Ringdown, quasinormal modes, and
  echoes}}, \href{https://doi.org/10.1103/PhysRevD.104.066021}{\emph{Phys. Rev.
  D} {\bfseries 104} (2021) 066021}
  [\href{https://arxiv.org/abs/2103.10960}{{\ttfamily 2103.10960}}].

\bibitem{Fucito:2023afe}
F.~Fucito and J.F.~Morales, \emph{{Post Newtonian emission of gravitational
  waves from binary systems: a gauge theory perspective}},
  \href{https://doi.org/10.1007/JHEP03(2024)106}{\emph{JHEP} {\bfseries 03}
  (2024) 106} [\href{https://arxiv.org/abs/2311.14637}{{\ttfamily
  2311.14637}}].

\bibitem{Bini:2023fiz}
D.~Bini, T.~Damour and A.~Geralico, \emph{{Comparing One-loop Gravitational
  Bremsstrahlung Amplitudes to the Multipolar-Post-Minkowskian Waveform}},
  \href{https://arxiv.org/abs/2309.14925}{{\ttfamily 2309.14925}}.

\bibitem{Bini:2024ijq}
D.~Bini, T.~Damour and A.~Geralico, \emph{{Gravitational Bremsstrahlung
  Waveform at the fourth Post-Minkowskian order and the second Post-Newtonian
  level}},  \href{https://arxiv.org/abs/2407.02076}{{\ttfamily 2407.02076}}.

\bibitem{Bini:2024rsy}
D.~Bini, T.~Damour, S.~De~Angelis, A.~Geralico, A.~Herderschee, R.~Roiban
  et~al., \emph{{Gravitational waveforms: A tale of two formalisms}},
  \href{https://doi.org/10.1103/PhysRevD.109.125008}{\emph{Phys. Rev. D}
  {\bfseries 109} (2024) 125008}
  [\href{https://arxiv.org/abs/2402.06604}{{\ttfamily 2402.06604}}].

\bibitem{Jakobsen:2021smu}
G.U.~Jakobsen, G.~Mogull, J.~Plefka and J.~Steinhoff, \emph{{Classical
  Gravitational Bremsstrahlung from a Worldline Quantum Field Theory}},
  \href{https://doi.org/10.1103/PhysRevLett.126.201103}{\emph{Phys. Rev. Lett.}
  {\bfseries 126} (2021) 201103}
  [\href{https://arxiv.org/abs/2101.12688}{{\ttfamily 2101.12688}}].

\bibitem{Mougiakakos:2021ckm}
S.~Mougiakakos, M.M.~Riva and F.~Vernizzi, \emph{{Gravitational Bremsstrahlung
  in the post-Minkowskian effective field theory}},
  \href{https://doi.org/10.1103/PhysRevD.104.024041}{\emph{Phys. Rev. D}
  {\bfseries 104} (2021) 024041}
  [\href{https://arxiv.org/abs/2102.08339}{{\ttfamily 2102.08339}}].

\bibitem{Cristofoli:2021vyo}
A.~Cristofoli, R.~Gonzo, D.A.~Kosower and D.~O'Connell, \emph{{Waveforms from
  amplitudes}}, \href{https://doi.org/10.1103/PhysRevD.106.056007}{\emph{Phys.
  Rev. D} {\bfseries 106} (2022) 056007}
  [\href{https://arxiv.org/abs/2107.10193}{{\ttfamily 2107.10193}}].

\bibitem{DiVecchia:2023frv}
P.~Di~Vecchia, C.~Heissenberg, R.~Russo and G.~Veneziano, \emph{{The
  gravitational eikonal: from particle, string and brane collisions to
  black-hole encounters}},  \href{https://arxiv.org/abs/2306.16488}{{\ttfamily
  2306.16488}}.

\bibitem{Brandhuber:2023hhy}
A.~Brandhuber, G.R.~Brown, G.~Chen, S.~De~Angelis, J.~Gowdy and G.~Travaglini,
  \emph{{One-loop gravitational bremsstrahlung and waveforms from a heavy-mass
  effective field theory}},
  \href{https://doi.org/10.1007/JHEP06(2023)048}{\emph{JHEP} {\bfseries 06}
  (2023) 048} [\href{https://arxiv.org/abs/2303.06111}{{\ttfamily
  2303.06111}}].

\bibitem{Herderschee:2023fxh}
A.~Herderschee, R.~Roiban and F.~Teng, \emph{{The sub-leading scattering
  waveform from amplitudes}},
  \href{https://doi.org/10.1007/JHEP06(2023)004}{\emph{JHEP} {\bfseries 06}
  (2023) 004} [\href{https://arxiv.org/abs/2303.06112}{{\ttfamily
  2303.06112}}].

\bibitem{Elkhidir:2023dco}
A.~Elkhidir, D.~O'Connell, M.~Sergola and I.A.~Vazquez-Holm, \emph{{Radiation
  and reaction at one loop}},
  \href{https://doi.org/10.1007/JHEP07(2024)272}{\emph{JHEP} {\bfseries 07}
  (2024) 272} [\href{https://arxiv.org/abs/2303.06211}{{\ttfamily
  2303.06211}}].

\bibitem{Georgoudis:2023lgf}
A.~Georgoudis, C.~Heissenberg and I.~Vazquez-Holm, \emph{{Inelastic
  exponentiation and classical gravitational scattering at one loop}},
  \href{https://doi.org/10.1007/JHEP06(2023)126}{\emph{JHEP} {\bfseries 2023}
  (2023) 126} [\href{https://arxiv.org/abs/2303.07006}{{\ttfamily
  2303.07006}}].

\bibitem{DeAngelis:2023lvf}
S.~De~Angelis, R.~Gonzo and P.P.~Novichkov, \emph{{Spinning waveforms from KMOC
  at leading order}},  \href{https://arxiv.org/abs/2309.17429}{{\ttfamily
  2309.17429}}.

\bibitem{Brandhuber:2023hhl}
A.~Brandhuber, G.R.~Brown, G.~Chen, J.~Gowdy and G.~Travaglini, \emph{{Resummed
  spinning waveforms from five-point amplitudes}},
  \href{https://arxiv.org/abs/2310.04405}{{\ttfamily 2310.04405}}.

\bibitem{Georgoudis:2023eke}
A.~Georgoudis, C.~Heissenberg and R.~Russo, \emph{{An eikonal-inspired approach
  to the gravitational scattering waveform}},
  \href{https://doi.org/10.1007/JHEP03(2024)089}{\emph{JHEP} {\bfseries 03}
  (2024) 089} [\href{https://arxiv.org/abs/2312.07452}{{\ttfamily
  2312.07452}}].

\bibitem{Georgoudis:2023ozp}
A.~Georgoudis, C.~Heissenberg and I.~Vazquez-Holm, \emph{{Addendum to:
  Inelastic exponentiation and classical gravitational scattering at one
  loop}}, \href{https://doi.org/10.1007/JHEP02(2024)161}{\emph{JHEP} {\bfseries
  2024} (2024) 161} [\href{https://arxiv.org/abs/2312.14710}{{\ttfamily
  2312.14710}}].

\bibitem{Georgoudis:2024pdz}
A.~Georgoudis, C.~Heissenberg and R.~Russo, \emph{{Post-Newtonian multipoles
  from the next-to-leading post-Minkowskian gravitational waveform}},
  \href{https://doi.org/10.1103/PhysRevD.109.106020}{\emph{Phys. Rev. D}
  {\bfseries 109} (2024) 106020}
  [\href{https://arxiv.org/abs/2402.06361}{{\ttfamily 2402.06361}}].

\bibitem{Brunello:2024ibk}
G.~Brunello and S.~De~Angelis, \emph{{An improved framework for computing
  waveforms}}, \href{https://doi.org/10.1007/JHEP07(2024)062}{\emph{JHEP}
  {\bfseries 07} (2024) 062}
  [\href{https://arxiv.org/abs/2403.08009}{{\ttfamily 2403.08009}}].

\bibitem{Alessio:2024onn}
F.~Alessio, P.~Di~Vecchia and C.~Heissenberg, \emph{{Logarithmic soft theorems
  and soft spectra}},  \href{https://arxiv.org/abs/2407.04128}{{\ttfamily
  2407.04128}}.

\bibitem{Fucito:2024wlg}
F.~Fucito, J.F.~Morales and R.~Russo, \emph{{Gravitational wave forms for
  extreme mass ratio collisions from supersymmetric gauge theories}},
  \href{https://arxiv.org/abs/2408.07329}{{\ttfamily 2408.07329}}.

\bibitem{Goldberger:2009qd}
W.D.~Goldberger and A.~Ross, \emph{{Gravitational radiative corrections from
  effective field theory}},
  \href{https://doi.org/10.1103/PhysRevD.81.124015}{\emph{Phys. Rev. D}
  {\bfseries 81} (2010) 124015}
  [\href{https://arxiv.org/abs/0912.4254}{{\ttfamily 0912.4254}}].

\bibitem{Trestini:2023wwg}
D.~Trestini and L.~Blanchet, \emph{{Gravitational-wave tails of memory}},
  \href{https://doi.org/10.1103/PhysRevD.107.104048}{\emph{Phys. Rev. D}
  {\bfseries 107} (2023) 104048}
  [\href{https://arxiv.org/abs/2301.09395}{{\ttfamily 2301.09395}}].

\bibitem{Almeida:2021jyt}
G.L.~Almeida, S.~Foffa and R.~Sturani, \emph{{Gravitational multipole
  renormalization}},
  \href{https://doi.org/10.1103/PhysRevD.104.084095}{\emph{Phys. Rev. D}
  {\bfseries 104} (2021) 084095}
  [\href{https://arxiv.org/abs/2107.02634}{{\ttfamily 2107.02634}}].

\bibitem{Edison:2023qvg}
A.~Edison and M.~Levi, \emph{{Higher-order tails and RG flows due to scattering
  of gravitational radiation from binary inspirals}},
  \href{https://doi.org/10.1007/JHEP08(2024)161}{\emph{JHEP} {\bfseries 08}
  (2024) 161} [\href{https://arxiv.org/abs/2310.20066}{{\ttfamily
  2310.20066}}].

\bibitem{Fioravanti:2021dce}
D.~Fioravanti and D.~Gregori, \emph{{A new method for exact results on
  Quasinormal Modes of Black Holes}},
  \href{https://arxiv.org/abs/2112.11434}{{\ttfamily 2112.11434}}.

\bibitem{Bianchi:2022wku}
M.~Bianchi and G.~Di~Russo, \emph{{Turning rotating D-branes and black holes
  inside out their photon-halo}},
  \href{https://doi.org/10.1103/PhysRevD.106.086009}{\emph{Phys. Rev. D}
  {\bfseries 106} (2022) 086009}
  [\href{https://arxiv.org/abs/2203.14900}{{\ttfamily 2203.14900}}].

\bibitem{Dodelson:2022eiz}
M.~Dodelson and A.~Zhiboedov, \emph{{Gravitational orbits, double-twist mirage,
  and many-body scars}},
  \href{https://doi.org/10.1007/JHEP12(2022)163}{\emph{JHEP} {\bfseries 12}
  (2022) 163} [\href{https://arxiv.org/abs/2204.09749}{{\ttfamily
  2204.09749}}].

\bibitem{Dodelson:2022yvn}
M.~Dodelson, A.~Grassi, C.~Iossa, D.~Panea~Lichtig and A.~Zhiboedov,
  \emph{{Holographic thermal correlators from supersymmetric instantons}},
  \href{https://doi.org/10.21468/SciPostPhys.14.5.116}{\emph{SciPost Phys.}
  {\bfseries 14} (2023) 116}
  [\href{https://arxiv.org/abs/2206.07720}{{\ttfamily 2206.07720}}].

\bibitem{Imaizumi:2022qbi}
K.~Imaizumi, \emph{{Quasi-normal modes for the D3-branes and Exact WKB
  analysis}}, \href{https://doi.org/10.1016/j.physletb.2022.137450}{\emph{Phys.
  Lett. B} {\bfseries 834} (2022) 137450}
  [\href{https://arxiv.org/abs/2207.09961}{{\ttfamily 2207.09961}}].

\bibitem{Fioravanti:2022bqf}
D.~Fioravanti, D.~Gregori and H.~Shu, \emph{{Integrability, susy $SU(2)$ matter
  gauge theories and black holes}},
  \href{https://arxiv.org/abs/2208.14031}{{\ttfamily 2208.14031}}.

\bibitem{Imaizumi:2022dgj}
K.~Imaizumi, \emph{{Exact conditions for quasi-normal modes of extremal
  M5-branes and exact WKB analysis}},
  \href{https://doi.org/10.1016/j.nuclphysb.2023.116221}{\emph{Nucl. Phys. B}
  {\bfseries 992} (2023) 116221}
  [\href{https://arxiv.org/abs/2212.04738}{{\ttfamily 2212.04738}}].

\bibitem{Bianchi:2022qph}
M.~Bianchi and G.~Di~Russo, \emph{{2-charge circular fuzz-balls and their
  perturbations}}, \href{https://doi.org/10.1007/JHEP08(2023)217}{\emph{JHEP}
  {\bfseries 08} (2023) 217}
  [\href{https://arxiv.org/abs/2212.07504}{{\ttfamily 2212.07504}}].

\bibitem{Dodelson:2023vrw}
M.~Dodelson, C.~Iossa, R.~Karlsson and A.~Zhiboedov, \emph{{A thermal product
  formula}}, \href{https://doi.org/10.1007/JHEP01(2024)036}{\emph{JHEP}
  {\bfseries 01} (2024) 036}
  [\href{https://arxiv.org/abs/2304.12339}{{\ttfamily 2304.12339}}].

\bibitem{Bianchi:2023rlt}
M.~Bianchi, C.~Di~Benedetto, G.~Di~Russo and G.~Sudano, \emph{{Charge
  instability of JMaRT geometries}},
  \href{https://doi.org/10.1007/JHEP09(2023)078}{\emph{JHEP} {\bfseries 09}
  (2023) 078} [\href{https://arxiv.org/abs/2305.00865}{{\ttfamily
  2305.00865}}].

\bibitem{Bianchi:2023sfs}
M.~Bianchi, G.~Di~Russo, A.~Grillo, J.F.~Morales and G.~Sudano, \emph{{On the
  stability and deformability of top stars}},
  \href{https://doi.org/10.1007/JHEP12(2023)121}{\emph{JHEP} {\bfseries 12}
  (2023) 121} [\href{https://arxiv.org/abs/2305.15105}{{\ttfamily
  2305.15105}}].

\bibitem{Giusto:2023awo}
S.~Giusto, C.~Iossa and R.~Russo, \emph{{The black hole behind the cut}},
  \href{https://doi.org/10.1007/JHEP10(2023)050}{\emph{JHEP} {\bfseries 10}
  (2023) 050} [\href{https://arxiv.org/abs/2306.15305}{{\ttfamily
  2306.15305}}].

\bibitem{BarraganAmado:2023apy}
J.~Barrag\'an~Amado, K.~Kwon and B.~Gwak, \emph{{Absorption cross section in
  gravity\textquoteright{}s rainbow from confluent Heun equation}},
  \href{https://doi.org/10.1088/1361-6382/ad1b92}{\emph{Class. Quant. Grav.}
  {\bfseries 41} (2024) 035005}
  [\href{https://arxiv.org/abs/2307.12824}{{\ttfamily 2307.12824}}].

\bibitem{Lei:2023mqx}
Y.~Lei, H.~Shu, K.~Zhang and R.-D.~Zhu, \emph{{Quasinormal modes of C-metric
  from SCFTs}}, \href{https://doi.org/10.1007/JHEP02(2024)140}{\emph{JHEP}
  {\bfseries 02} (2024) 140}
  [\href{https://arxiv.org/abs/2308.16677}{{\ttfamily 2308.16677}}].

\bibitem{BarraganAmado:2023wxt}
J.~Barrag\'an~Amado and B.~Gwak, \emph{{Scalar quasi-normal modes of
  accelerating Kerr-Newman-AdS black holes}},
  \href{https://doi.org/10.1007/JHEP02(2024)189}{\emph{JHEP} {\bfseries 2024}
  (2024) 189} [\href{https://arxiv.org/abs/2309.11355}{{\ttfamily
  2309.11355}}].

\bibitem{Ge:2024jdx}
X.-H.~Ge, M.~Matsumoto and K.~Zhang, \emph{{Duality between Seiberg-Witten
  theory and black hole superradiance}},
  \href{https://doi.org/10.1007/JHEP05(2024)336}{\emph{JHEP} {\bfseries 05}
  (2024) 336} [\href{https://arxiv.org/abs/2402.17441}{{\ttfamily
  2402.17441}}].

\bibitem{Cipriani:2024ygw}
A.~Cipriani, C.~Di~Benedetto, G.~Di~Russo, A.~Grillo and G.~Sudano,
  \emph{{Charge (in)stability and superradiance of Topological Stars}},
  \href{https://doi.org/10.1007/JHEP07(2024)143}{\emph{JHEP} {\bfseries 07}
  (2024) 143} [\href{https://arxiv.org/abs/2405.06566}{{\ttfamily
  2405.06566}}].

\bibitem{Arnaudo:2024rhv}
P.~Arnaudo, G.~Bonelli and A.~Tanzini, \emph{{One loop effective actions in
  Kerr-(A)dS black holes}},
  \href{https://doi.org/10.1103/PhysRevD.110.106006}{\emph{Phys. Rev. D}
  {\bfseries 110} (2024) 106006}
  [\href{https://arxiv.org/abs/2405.13830}{{\ttfamily 2405.13830}}].

\bibitem{Bianchi:2024vmi}
M.~Bianchi, D.~Bini and G.~Di~Russo, \emph{{Scalar perturbations of
  topological-star spacetimes}},
  \href{https://doi.org/10.1103/PhysRevD.110.084077}{\emph{Phys. Rev. D}
  {\bfseries 110} (2024) 084077}
  [\href{https://arxiv.org/abs/2407.10868}{{\ttfamily 2407.10868}}].

\bibitem{Bianchi:2024mlq}
M.~Bianchi, G.~Dibitetto and J.F.~Morales, \emph{{Gauge theory meets
  cosmology}}, \href{https://doi.org/10.1088/1475-7516/2024/12/040}{\emph{JCAP}
  {\bfseries 12} (2024) 040}
  [\href{https://arxiv.org/abs/2408.03243}{{\ttfamily 2408.03243}}].

\bibitem{DiRusso:2024hmd}
G.~Di~Russo, F.~Fucito and J.F.~Morales, \emph{{Tidal resonances for
  fuzzballs}}, \href{https://doi.org/10.1007/JHEP04(2024)149}{\emph{JHEP}
  {\bfseries 04} (2024) 149}
  [\href{https://arxiv.org/abs/2402.06621}{{\ttfamily 2402.06621}}].

\bibitem{Bianchi:2024rod}
M.~Bianchi, D.~Bini and G.~Di~Russo, \emph{{Scalar waves in a Topological Star
  spacetime: self-force and radiative losses}},
  \href{https://arxiv.org/abs/2411.19612}{{\ttfamily 2411.19612}}.

\bibitem{Aminov:2024aan}
G.~Aminov and P.~Arnaudo, \emph{{Basics of Multiple Polyexponential
  Integrals}},  \href{https://arxiv.org/abs/2409.06760}{{\ttfamily
  2409.06760}}.

\bibitem{BarraganAmado:2024tfu}
J.~Barrag\'an~Amado, S.~Chakrabortty and A.~Maurya, \emph{{The effect of
  resummation on retarded Green\textquoteright{}s function and greybody factor
  in AdS black holes}},
  \href{https://doi.org/10.1007/JHEP11(2024)070}{\emph{JHEP} {\bfseries 11}
  (2024) 070} [\href{https://arxiv.org/abs/2409.07370}{{\ttfamily
  2409.07370}}].

\bibitem{Bautista:2024emt}
Y.F.~Bautista, Y.-T.~Huang and J.-W.~Kim, \emph{{Absorptive effects in black
  hole scattering}},
  \href{https://doi.org/10.1103/PhysRevD.111.044043}{\emph{Phys. Rev. D}
  {\bfseries 111} (2025) 044043}
  [\href{https://arxiv.org/abs/2411.03382}{{\ttfamily 2411.03382}}].

\bibitem{Liu:2024eut}
P.~Liu and R.-D.~Zhu, \emph{{Notes on Quasinormal Modes of charged de Sitter
  Blackholes from Quiver Gauge Theories}},
  \href{https://arxiv.org/abs/2412.18359}{{\ttfamily 2412.18359}}.

\bibitem{Silva:2025khf}
H.O.~Silva, J.-W.~Kim and M.V.S.~Saketh, \emph{{Kerr-Newman quasinormal modes
  and Seiberg-Witten theory}},
  \href{https://doi.org/10.1103/PhysRevD.111.104021}{\emph{Phys. Rev. D}
  {\bfseries 111} (2025) 104021}
  [\href{https://arxiv.org/abs/2502.17488}{{\ttfamily 2502.17488}}].

\bibitem{Aprile:2025hlt}
F.~Aprile, S.~Giusto and R.~Russo, \emph{{Four-point correlators with BPS bound
  states in AdS$_3$ and AdS$_5$}},
  \href{https://arxiv.org/abs/2503.02855}{{\ttfamily 2503.02855}}.

\bibitem{Bhatta:2025kil}
A.~Bhatta and A.~Maurya, \emph{{Greybody Factor in AdS Black Brane
  Spectroscopy: A Study with Scalar and Electromagnetic Perturbations}},
  \href{https://arxiv.org/abs/2503.22805}{{\ttfamily 2503.22805}}.

\bibitem{BHPToolkit}
``{Black Hole Perturbation Toolkit}.''
  (\href{http://bhptoolkit.org/}{bhptoolkit.org}).

\bibitem{Nekrasov:2009rc}
N.A.~Nekrasov and S.L.~Shatashvili, \emph{{Quantization of Integrable Systems
  and Four Dimensional Gauge Theories}},  in \emph{{16th International Congress
  on Mathematical Physics}}, pp.~265--289, 2010,
  \href{https://doi.org/10.1142/9789814304634_0015}{DOI}
  [\href{https://arxiv.org/abs/0908.4052}{{\ttfamily 0908.4052}}].

\bibitem{Newman:1961qr}
E.~Newman and R.~Penrose, \emph{{An Approach to gravitational radiation by a
  method of spin coefficients}},
  \href{https://doi.org/10.1063/1.1724257}{\emph{J. Math. Phys.} {\bfseries 3}
  (1962) 566}.

\bibitem{Poghosyan:2020zzg}
H.~Poghosyan, \emph{{Recursion relation for instanton counting for SU(2) $
  \mathcal{N} $ = 2 SYM in NS limit of $\Omega$ background}},
  \href{https://doi.org/10.1007/JHEP05(2021)088}{\emph{JHEP} {\bfseries 05}
  (2021) 088} [\href{https://arxiv.org/abs/2010.08498}{{\ttfamily
  2010.08498}}].

\end{thebibliography}
\end{document}